\documentclass{article}
\usepackage[utf8]{inputenc}

\usepackage{lipsum}
\usepackage{url}

\usepackage[T1]{fontenc}
\usepackage[utf8]{inputenc}
\usepackage[english]{babel}
\usepackage{times}
\usepackage[left=2cm, right=2cm, top=3cm,bottom=3cm]{geometry}
\usepackage{amsmath}
\usepackage{amssymb}
\usepackage{amsthm}
\usepackage{hyperref}
\usepackage{graphicx}
\usepackage{siunitx}
\usepackage{subcaption}
\usepackage{caption}
\usepackage{xcolor}
\usepackage{hyperref}
\usepackage{longtable}
\usepackage{verbatimbox}

\usepackage[title]{appendix}
\usepackage{bm}
\usepackage{empheq}
\usepackage{placeins}

\usepackage{xcolor}
\hypersetup{
colorlinks,
linkcolor={red!50!black},
citecolor={blue!50!black},
urlcolor={blue!80!black}
}

\usepackage{amsmath}
\usepackage{amssymb}
\usepackage{amsthm}
\usepackage{hyperref}
\usepackage{graphicx}
\usepackage{subcaption}
\usepackage{caption}
\usepackage{xcolor}

\usepackage{longtable}
\usepackage{verbatimbox}
\usepackage[title]{appendix}
\usepackage{bm}
\usepackage{empheq}
\usepackage{placeins}
\theoremstyle{plain}
\newtheorem{theorem}{Theorem}[section]

\newtheorem{remark}[theorem]{Remark}

\theoremstyle{definition}
\theoremstyle{remark}
\numberwithin{equation}{section}

\theoremstyle{plain} 
\usepackage{mathtools}
\usepackage{bm}
\usepackage{pgfplots}
\pgfplotsset{/pgf/number format/use comma,compat=newest}

\renewcommand\epsilon{\varepsilon}

\newcommand{\tens}[1]{\mathsf{#1}}

\usepackage{subcaption}
\usepackage[small, sc]{titlesec}
\usepackage{array}

\usepackage{xcolor}
\hypersetup{
colorlinks,
linkcolor={red!50!black},
citecolor={blue!50!black},
urlcolor={blue!80!black}
}
\usepackage{booktabs}
\usepackage[
backend=bibtex,
style=numeric,
sorting=nty,
maxbibnames=50,
giveninits=true
]{biblatex}
\usepackage{csquotes}
\renewbibmacro{in:}{%
 \ifentrytype{article}{}{}}
\DeclareFieldFormat[article]{volume}{\textbf{#1}}
\DeclareFieldFormat[article]{title}{\textit{#1}}
\DeclareFieldFormat[article]{journaltitle}{#1}
\DeclareFieldFormat[article]{number}{\textbf{#1}}
\renewbibmacro*{volume+number+eid}{%
  \printfield{volume}%
\setunit*{\textbf{\addcolon}}%<---- was \setunit*{\adddot}%
  \printfield{number}\setunit{\addcomma\space}%
  \printfield{eid}}

\begin{document}

\title{\textsc{
Wrinkling of a bilayered cylinder under tension
}
}

\author{ \textsc{C.\ Lonati}$^1$%\thanks{\href{mailto:chiara.lonati@polito.it}{\texttt{chiara.lonati@polito.it}}}
 ,\,
\textsc{S.P. Venkata}$^{2,3,4}$\thanks{\href{ sairampv44@gmail.com}{
\texttt{ sairampv44@gmail.com}}} ,\,\textsc{K.\ Moerman}$^5$
 ,\,\textsc{O. Petrenko}$^2$
 ,\, \textsc{M. Destrade}$^{2,3}$
 ,\, \textsc{V. Balbi}$^2$
\bigskip\\
\normalsize$^1$ mathLab, Mathematics Area, SISSA - International School for Advanced Studies,  Trieste, Italy. \\
\normalsize$^2$ School of Mathematical and Statistical Sciences, University of Galway, University Road, Galway, Ireland. \\
\normalsize$^3$ School of Materials and Mechanical Engineering, University College Dublin, Belfield, Dublin 4, Ireland. \\
\normalsize$^4$ Biomedical Design, School of Mechanical Engineering, Technological University Dublin, Dublin 24, Ireland. \\
\normalsize$^5$ School of Engineering, University of Galway, University Road, Galway, Ireland. 
}
\date{}

\maketitle
\begin{abstract}
We study near-surface tension instabilities in bilayered cylinders composed of an incompressible core coated with a thin compressible film. 
In theory, circumferential wrinkles with axial wavefronts may appear on the outer surface when the cylinder is stretched. 
We analyze these instabilities through modelling, numerics, finite element simulations, and controlled experiments. 
We show that the wrinkle amplitude and wavelength depend on the Poisson ratio of the coating and the ratio of the film and substrate shear moduli. 
We find that the lower the film's Poisson ratio, the earlier the wrinkles emerge under tension.
This observation motivates us to use auxetic films (with negative Poisson ratios) with bespoke microstructures in the finite element simulations and the experiments.
For the latter, we fabricated 3D-printed soft cylinders with stiff auxetic outer layers and subjected them to axial loading. 
Although simulations predicted clear wrinkle formation, experimental observation proved more elusive due to the small amplitude of the wrinkles.
\end{abstract}
\medskip


   \section{Introduction}
\label{5.1}

%%%%%%%%%%%%%%%%%%%%%%%%%%%%%

Multilayered structures are ubiquitous in nature, particularly within biological systems where functional behaviour often emerges from the intricate interplay of geometry and material composition. 
These architectures are typically composed of a soft, thick, compliant substrate coated with a thin, stiffer outer layer, enabling complex morphological responses under physiological loading. 
Notable examples include the cerebral cortex, where differential growth between layers causes characteristic folding patterns \cite{riccobelli2020surface,balbi2020mechanics}; the dermis-epidermis structure of the skin, which accommodates large deformations while preserving its protective role \cite{ciarletta2013shear};  tubular structures such as axons \cite{PRE_2021,andrini2022mathematical} and the gastrointestinal tract \cite{balbi2013morpho,balbi2015morphoelastic}, where layered walls enhance both mechanical integrity and dynamic motion. 

% These biological architectures suggest that regulated mechanical instabilities play a fundamental role in shaping tissues, generating surface patterns, and guiding their functional organization.

When layered elastic structures are subjected to large deformations such as axial tension, compression, shear, bending, torsion, etc., surface instabilities often develop in the form of wrinkle patterns. These patterns are influenced by the geometry of the structure, the nature of the applied loads, the direction of the loading, and the material properties of the individual layers. 
Wrinkling phenomena in soft multilayers are common in nature and have also found applications in flexible electronics \cite{bowden1998}, soft robotics \cite{genzer2006}, stretchable optics \cite{Jiang2007}, and biomedical devices \cite{kim2011}.

Wrinkling in planar bilayers has been extensively examined, beginning with the foundational work of Biot on surface instability \cite{biot1963surface} and later, in the context of stiff films on compliant substrates by, among others, Cai and Fu \cite{cai1999imperfection}, Cao and Hutchinson \cite{bilayercao} and Hohlfeld and Mahadevan \cite{Hohlfield2011}. 
In cylindrical systems, Wilkes provided early insight into the buckling behaviour of compressed tubes \cite{wilkes1955stability}, while subsequent theoretical and numerical studies have explored bifurcations in confined and prestressed cylindrical configurations \cite{bigoni2001bifurcations, curatolo2023circumferential}. 
In addition, instabilities in spherical geometries have been examined within biological and morphoelastic contexts \cite{riccobelli2017shape}.
For incompressible solids, studies have also examined the role of growth \cite{moulton2011circumferential, balbi2013morpho}, residual stress \cite{riccobelli2017shape}, material anisotropy \cite{liu2021influence, destrade2015incremental}, shear \cite{destrade2008surface,ciarletta2013shear}, torsion \cite{ciarletta2014torsion}, bending \cite{destrade2009bending}, pre-stretching \cite{balbi2015helical, zhao2021stability}, etc., and combinations thereof \cite{du2019influence}. 

Compared to these extensively studied incompressible systems, only limited research has considered compressible films, despite their importance in applications including polymer foams \cite{GibsonAshby1997} used in protective padding and thermal insulation, dry gels \cite{Maeda2008} employed in soft actuators, and architectured materials with tunable compressibility or auxetic behaviour \cite{katia2010}. 
These materials can undergo considerable volume changes under mechanical loading and exhibit surface instability patterns, an aspect which remains relatively unexplored, particularly under axial tension. 
Although some theoretical works have been conducted to address the role of compressibility in elastic stability \cite{obrezkov2016equilibrium}, significant gaps remain in our understanding of how these materials wrinkle in multilayered or curved configurations.

Compressible hyperelastic models such as the Blatz–Ko model \cite{parl1962application} provide a well-established framework for capturing nonlinear behaviours in isotropic compressible solids. 
However, studies on wrinkling in these materials are still limited, especially in cylindrical geometries. 
Although analytical solutions are available for certain idealised compressible problems \cite{carroll1990finite, carroll2001some, carroll2005compressible}, numerical methods are typically required to resolve both the base deformation and the onset of instability in more general or applied settings.

A recent study by Venkata et al. \cite{venkata2025wrinkling} highlighted the role of compressibility in the wrinkling of rectangular bilayers subjected to uniaxial tension. 
Their analysis, which focused on auxetic films bonded to nearly incompressible substrates using the Blatz-Ko formulation, showed that wrinkle orientation and onset are governed by both the Poisson ratios and the shear moduli contrast. 
In particular, it was shown that wrinkles could form with wavefronts parallel to the loading direction, with critical stretch values decreasing as the coating became more compressible. 
These findings motivate the current work, which extends the investigation to cylindrical geometries relevant to biological tissues such as arteries and neural structures.

While axial compression-induced wrinkling in cylinders has been well documented through theoretical, numerical, and experimental investigations \cite{Zhao2014, Zhan2022, kordolemis}, tension-induced instabilities remain relatively underexplored in theory. However, recent experiments on swelling-induced growth in tubular bilayers have revealed wrinkling patterns consistent with tension-driven surface instabilities \cite{liu2022experimental}. This work aims to provide a detailed characterization of such tension-driven modes, emphasizing the roles of material compressibility and moduli contrast.

Hence, we examine the onset of circumferential wrinkling in cylindrical bilayers composed of an incompressible core and a compressible, thin coating under uniaxial tension. 
We analyze how the critical stretch ratio and associated wavenumber depend on the Poisson ratio of the coating and the shear modulus mismatch between the two layers. 
Our approach is to use the theory of incremental elasticity \cite{ogden1997non, ogden2001elements} to investigate the onset of bifurcation around a finitely deformed configuration. 
We cast the associated governing equations into the Stroh form and integrate them numerically with the Riccati transfer matrix method, which is particularly efficient in cylindrical geometries \cite{destrade2009bending, Shuvalov2010, destrade2010deforming}.
Our results show that a high contrast in Poisson’s ratio, particularly when the coating is auxetic and the substrate near-incompressible,  combined with a significant shear modulus mismatch, can promote the onset of instability at relatively low stretch values and generate distinct wrinkling patterns.

The structure of the paper is as follows. In Section~\ref{5.2}, we introduce the governing equations for the bilayered cylinder. Section~\ref{basic} presents the base solution for the finitely deformed configuration, and Section~\ref{5.4} derives the incremental equations using the Stroh formulation, together with the differential Riccati equation used to integrate them numerically. Section~\ref{5.5} describes the construction of the corresponding Finite Element models. Finally, Section~\ref{5.6} presents our results: validation of the semi-analytical model against the Finite Element simulations, the influence of the coating's Poisson ratio and shear modulus contrast on the critical stretch and wavenumber, and preliminary experiments on 3D-printed bilayered cylinders with architectured auxetic coatings, whose fabrication and mechanical testing are detailed in the appendix.

%%%%%%%%%%%%%%%%%%%%%%
\section{Modelling}
\label{5.2}

%%%%%%%%%%%%%%%%%%%%%%%

%+++++++++++++++++++++++++

\subsection{Kinematics}

%++++++++++++++++++++++++++

We call $\mathcal B_0$ the reference configuration and $\mathcal B$ the current configuration, and use cylindrical coordinates for the positions $\textbf{X}=(R,\Theta,Z)$ in $\mathcal B_0$ and $\textbf{x}(r,\theta,z)$ in $\mathcal B$.
The cylinder is made up of two perfectly bonded layers: an inner incompressible substrate with initial radius $R_m$ and an outer compressible thin film with external initial radius $R_o$. 
Their common length, $L$ in $\mathcal B_0$ and $\ell$ in $\mathcal B$, is assumed large compared to their radii, so that end boundary effects can be neglected.

\begin{figure}[t]
    \centering
    \includegraphics[width=0.97\linewidth]{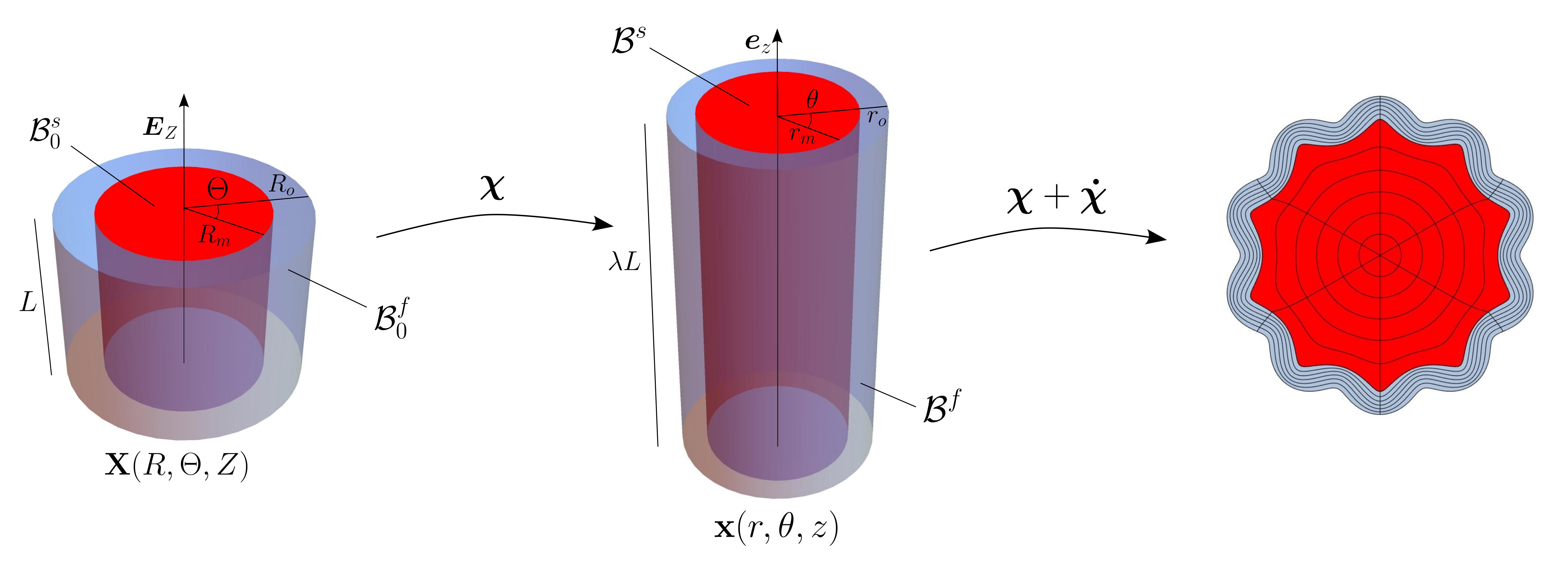}
    \caption{(left) The undeformed cylinder in its initial configuration $\mathcal B_0$ with initial height $L$. The substrate (in red) has initial radius $R_m$, and the film (in grey) has initial outer radius $R_o$. (centre) The stretched cylinder in its deformed configuration $\mathcal B$ with height $\ell=\lambda L$. The substrate has final radius $r_m$, and the film has final outer radius $r_o$. (right) Top view of the cylinder in the perturbed configuration, once circumferential wrinkling has occurred.}
    \label{cilindrofinale}
\end{figure}

The double-layer cylinder undergoes a uni-axial  extension, 
\begin{equation}\label{deformation}
     r^i=r^i(R), \qquad \theta=\Theta, \qquad z =\lambda\, Z, 
     \end{equation}
where $i=s,f$ refers to the substrate or the film, and $\lambda = \ell/L$. 
This large deformation $\mathbf x = \boldsymbol \chi(\mathbf X)$ is then incremented and we look for linearised solutions $\boldsymbol{\dot\chi}$ in its neighbourhood, in the shape of circumferential wrinkles with axial wavefronts, indicating the onset of linearised instability, see Figure \ref{cilindrofinale}.  

%++++++++++++++++++++++++++++++++++++++++++

\subsection{Constitutive assumptions and equilibrium equations}

%++++++++++++++++++++++++++++++++++++++++++

We assume that both constituents are homogeneous, hyperelastic and isotropic, so that we can define the strain energy functions $W^i(I_1,I_2,I_3)$ where $I_1$, $I_2$, $I_3$ are the first three principal invariants of the Cauchy-Green deformation tensors. 
We take the following form,
\begin{equation}\label{energy}
    W^i =\frac{\mu_i}{2}\left\{\alpha_i \left[I_1 - 3 + \frac{1-2\nu_i}{\nu_i}\left(I_3^{\frac{\nu_i}{2 \nu_i-1}}-1\right)\right]+(1-\alpha_i) \left[\frac{I_2}{I_3} - 3 + \frac{1-2\nu_i}{\nu_i}\left(I_3^{\frac{\nu_i}{2 \nu_i-1}} - 1\right)\right]\right\},
\end{equation}
which combines the standard Mooney-Rivlin form with the Blatz-Ko compressible form. Here $\mu_i$ is the initial shear modulus, $\alpha_i$ is the weighting Mooney-Rivlin parameter ($0\le\alpha_i\le1$) and $\nu_i$ is the Poisson ratio ($-1 \le \nu_i \le 0.5$).
The substrate is incompressible, so that $I_3 \equiv 1$ and $\nu_s=\tfrac 1 2$ for $W^s$.
% We also assume to deal with homogenous and isotropic materials, so that it will be $W=W(I_1,I_2,I_3)$ with $I_1=\operatorname{tr}\tens B$, $I_2=(1/2)((\operatorname{tr}\tens B)^2-\operatorname{tr}\tens B^2)$ and $I_3=\det \tens F$, where $\tens B^2$ naturally means the multiplication of $\tens B$ with itself.\\
% The inner substrate is made up of a Mooney-Rivlin incompressible material, with 
% $$W^\text s=\frac{\mu_s}{2}\left(\alpha_s (I_1^\text s-3)+(1-\alpha_s) (I_2^\text s-3)\right)$$
% where $\alpha_s$ and $\mu_s$ are material parameters of the substrate and in particular $\mu_s$ is the shear modulus. Being an incompressible material, we require $\det \tens{F}^\text s=1$.\\
% The outer thin coating is made up of a Blatz-Ko compressible material, with 
% $$W^c=\frac{\mu_c}{2}\left(\alpha_c \left(I_1^c-3+\frac{1}{\beta}\left((I_3^c)^{-\beta}-1\right)\right)+(1-\alpha_c) \left(\frac{I_2^c}{I_3^c}-3+\frac{1}{\beta}\left((I_3^c)^{-\beta}-1\right)\right)\right)$$
% where $\beta$, $\alpha_c$ and $\mu_c$ are material parameters of the coating and in particular $\mu_c$ is the shear modulus, while we have
% $$\beta =\frac{\nu }{1-2 \nu }$$ where $\nu$ is the Poisson ratio's of the film, $-1<\nu\leq 1/2$. However, the assumption of compressibility restricts the Poisson's ratio to the interval $-1<\nu<1/2$.\\

We can then calculate the Cauchy stress $\tens T$ in the substrate and film as 
% can be computed for the two layers.
% In the substrate, we have to enforce the incompressibility constraint using a Lagrange multiplier $p$, then we have  
% \begin{equation}
% \label{piolainc}
%     \tens{P}^\text s=\tens{T}^\text s({\tens{F}^\text s}^T)^{-1},\quad \tens{T}^\text s=\tens{F^\text s}\frac{\partial W^\text s}{\partial \tens{F}^\text s}-p \tens{I},\quad \text{in components}\quad \tens{T}_{ij}^\text s =\tens{F}^\text s_{ik}\frac{\partial W^\text s}{\partial \tens{F}^\text s_{jk}}-p\delta_{ij}.
% \end{equation}
% In the coating, the nominal and Cauchy stress tensor read 
\begin{equation}\label{stresses}
\tens{T}^s = \tens{F} \frac{\partial W^s}{\partial \tens{F}} - p\,\tens{I}, \qquad
 \tens{T}^f = \frac{1}{\det \tens{F}}\tens{F}^f\frac{\partial W^f}{\partial \tens{F}},
\end{equation}
% ,\quad \text{ in components }\quad\tens{T}_{ij}^\text n =\frac{1}{\det \tens{F}^\text n}\tens{F}^\text n_{ik}\frac{\partial W^\text n}{\partial \tens{F}^\text n_{jk}}.
% In addition to the perfect bonding on the layers, we introduce some mechanical assumptions on the system: the layers are equilibrium, the central axis of the cylinder is fixed, body forces vanish and radial stress is continuous at the interface and vanishing on the outer surface.\\
% From classical Continuum Mechanics, equilibrium equations for the linear momentum in the two layers can be written as 
where $\tens F = \partial \boldsymbol \chi/\partial \mathbf X$ is the deformation gradient and $p$ is the Lagrange multiplier required to enforce incompressibility in the substrate. 

The equilibrium equations are
\begin{equation}\label{equieqs}
    \operatorname{div} \tens{T}^i=\boldsymbol{0},\qquad i=\{ s,f\},
\end{equation}and we solve them subject to the following boundary conditions (bonded interface, traction-free outer surface),
\begin{equation}\label{bceqs}
r^s(R_m)=r^f(R_m), \qquad T^s_{rr}(r_m)=T^f_{rr}(r_m), \qquad T^f_{rr}(r_o)=0.    
\end{equation}
% with $\boldsymbol{N}$ and $\boldsymbol{n}$ the normal vectors at the interface in the reference and
% actual configurations respectively, i.e. $\boldsymbol{E}_R$ and $\boldsymbol{e}_r$ respectively.
%In the next section, we will show how to solve the equilibrium equations in \eqref{equieqs} by using the boundary conditions in \eqref{bceqs} and the continuity conditions in \eqref{conteq1} and \eqref{conteq2}. We will obtain the basic solution and perform a linear stability analysis to investigate the onset of wrinkling in the double-layered cylinder in tension.

%++++++++++++++++++++++++++++++++++++

%\subsection{Incremental stability analysis}\label{5.3}

%+++++++++++++++++++++++++++++++++++

%To study the onset of wrinkling, we use the method of incremental deformations superimposed on finite
%deformations (small-on-large method). The main idea is to first solve the equilibrium problem and find the basic solution. Then we perturb the basic solution by adding a small incremental displacement and we solve the linearised perturbed problem. This approach allows us to find a bifurcation condition which gives us the critical stretch and the wavenumber associated to the instability pattern (wrinkling).

%++++++++++++++++++++++++++++++++++++++++

\subsection{Large axial extension}\label{basic}

%++++++++++++++++++++++++++++++++++++++++

From \eqref{deformation}, we calculate the deformation gradient for each layer as
% We recover here an axisymmetric solution for the problem defined in the preceding section.
% We assume a
% base axisymmetric deformation in the form
% $$
% \begin{cases}
%     r=r(R),\\
%     \theta=\Theta,\\
%     z=\lambda Z,
% \end{cases}
% $$
% for both layers.

% The axisymmetric solution for the problem described above turn out to be analytic in the inner cylinder, but must be solved numerically in the compressible film.

% The deformation gradient for this defomration in cylindrical coordinates is the diagonal matrix
\begin{equation}\label{deformationgen}
\tens{F}=\operatorname{Diag}\left(\text{d}r/\text{d}R,r/R, \lambda\right).    
\end{equation}
% Alternatively, we can write:
% \begin{equation} 
% \tens{F}^\text s=\operatorname{Diag}( \lambda_r^\text s, \lambda_\theta^\text s ,\lambda_z^\text s),\qquad \tens{F}^\text f=\operatorname{Diag}(
%     \lambda_r^c,\lambda_\theta^c ,\lambda_z^c).
% \end{equation}
% where the following restrictions apply:
% \begin{equation}
% \lambda_z^\text s=\lambda_z^\text f=\lambda,
% \quad \lambda_r^\text s\lambda_\theta^\text s\lambda_z^\text s=1\quad \text{ and }\quad 
% \lambda_\theta^\text s=\lambda_\theta^\text f\,\,\text{ at the interface}.    
% \end{equation}
% To express that the two materials are subject to the same uni-axial tension along the axis, that the substrate is incompressible and the continuity of angular stretch at the interface (so that the cylinder is not deformed into an elliptical shaped cross-section), it must hold

In the inner layer, we solve the incompressibility condition $\text{det}\,\tens{F}^s= \lambda r'r/R=1$ with $r(0)=0$ to give $r(R)=R/\sqrt{\lambda}$ and 
\begin{equation}\label{defin}
    \tens{F}=\operatorname{Diag}\left(1/\sqrt{\lambda},1/\sqrt{\lambda},\lambda\right),
\end{equation}
showing that the substrate deforms homogeneously and the equation of equilibrium is automatically satisfied there.
The principal components of the Cauchy stress in the substrate follow as
% $$
% \scalemath{0.85}{
% \tens{P}^\text s=\operatorname{Diag}\left(\sqrt{\lambda}\left(\frac{\mu_s(1+\lambda^3+\alpha_s(-1+\lambda-\lambda^3))}{\lambda^2}-p\right),\sqrt{\lambda}\left(\frac{\mu_s(1+\lambda^3+\alpha_s(-1+\lambda-\lambda^3))}{\lambda^2}-p\right), (2+\alpha_s(\lambda-2))\mu_s-\frac{p}{\lambda}\right)}
% $$
\begin{equation}
T_{rr}^s=T_{\theta\theta}^s=\mu_s[\lambda + \lambda^{-2} + \alpha_s(\lambda^{-1} - \lambda -\lambda^{-2})] -p, \qquad  
T_{zz}^s= \mu_s[2 + \alpha_s(\lambda-2)]-p,
\end{equation}
% $$
% \scalemath{1}{
% \tens{T}^\text s=\operatorname{Diag}\left(\frac{\mu_s(1+\lambda^3+\alpha_s(-1+\lambda-\lambda^3))}{\lambda^2}-p,\frac{\mu_s(1+\lambda^3+\alpha_s(-1+\lambda-\lambda^3))}{\lambda^2}-,(2\!+\!\alpha_s(\lambda-2))\lambda\mu_s\!-\!p
% \right)}
% $$
where $p$ is to be determined once the equilibrium problem is solved in the film, by imposing the continuity conditions \eqref{bceqs}.
% Being the deformation homogeneous, $\tens{P}^\text s_{rr}=\tens{P}^\text s_{\theta\theta}$ and $\tens{T}^\text s_{rr}=\tens{T}^\text s_{\theta\theta}$, and the stresses are constant with respect to $R$. 
% The equilibrium equations $\operatorname{div} \tens{P}^\text s=\boldsymbol{0}$ and $\operatorname{div} \tens{T}^\text s=\boldsymbol{0}$ that reduce to 
% $$
% \frac{d\tens{T}_{rr}^\text s}{dr}+\frac{\tens{T}_{rr}^\text s-\tens{T}_{\theta\theta}^\text s}{r}=0
% $$
% are then automatically satisfied.

In the compressible film we determine the radial component of the deformation $r^f(R)$ by  solving numerically the equilibrium equations \eqref{equieqs}, together with the initial and continuity conditions \eqref{bceqs}, in Mathematica using the \texttt{NDSolve} function. 

% to find $r^c(R)$, $R\in [R_m,R_o]$. The complete expressions of $\tens{P}^c$ and $\tens{T}^c$ and the complete expressions for the system are reported in Appendices A.1 and A.2.\\

\begin{figure}[htbp]
    \centering
    \includegraphics[width=0.7\textwidth]{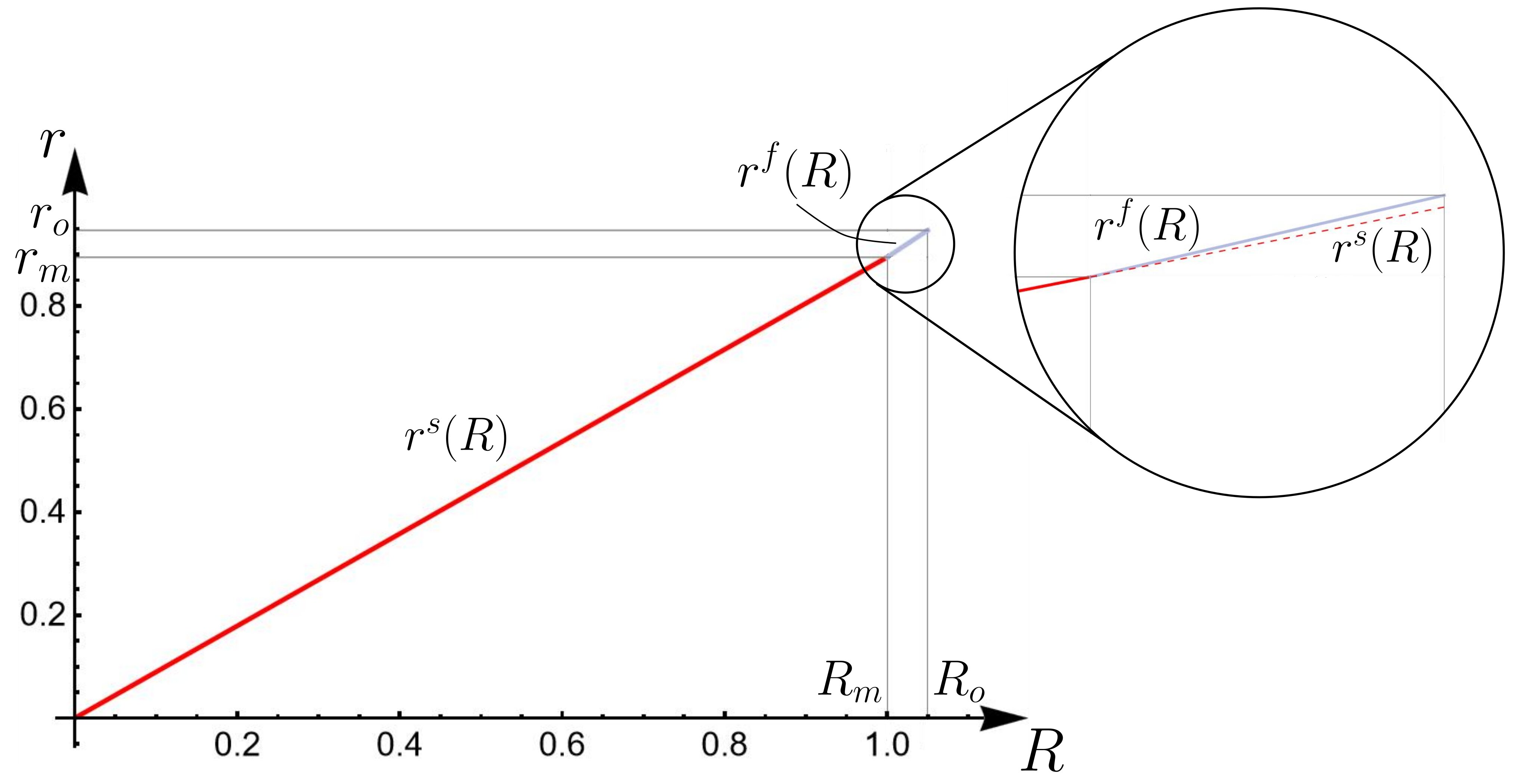}
    \caption{Radial solution in the deformed bilayered cylinder. In the core incompressible substrate, the homogeneous solution $r^s(R)$ is shown in red for $R\in[0,R_m]$ and  in the film, the numerical solution $r^f(R)$ is shown in grey for $R\in[R_m,R_o]$. The zoomed inset shows the deviation of the numerical solution from $r^s(R)$. The following representative parameters were chosen for this example: $\lambda=1.25$, $R_m=1$, $R_o=1.05$, $\nu_f=-0.5$, $\mu_f/\mu_s=10$, $\alpha_s=0.2$, $\alpha_f=0.5$.}
    \label{soluzioner}
\end{figure}

% In addition to this, the condition $\lim_{R\to R_m^-}{\tens{P}^\text s}\boldsymbol{N}=\lim_{R\to R_m^+}{\tens{P}^c}\boldsymbol{N}$ allows us to determine the value of $p$, whose expression is
% $$
% \begin{aligned}
%     p&=\frac{\mu_s \left(\alpha_s \left(-\lambda ^3+\lambda -1\right)+\lambda ^3+1\right)}{\lambda^2}-\frac{\mu_f \left(\alpha_f-\alpha_f {r^c}'(R)^2 \left(\frac{\lambda ^2 {r^c}(R)^2 {r^c}'(R)^2}{R^2}\right)^{-\beta }\right)}{\sqrt{\lambda } {r^c}'(R)^3}\\
%     &+\frac{\mu_f\left((\alpha_f-1) {r^c}'(R)^2 \left(\frac{\lambda ^2 {r^c}(R)^2 {r^c}'(R)^2}{R^2}\right)^{\beta }+\alpha_f {r^c}'(R)^4-1\right)}{\sqrt{\lambda } {r^c}'(R)^3}.
% \end{aligned}
% $$
We plot the corresponding radial particle positions $r^s(R)$ and $r^f(R)$ in the deformed cylinder for a representative set of material and geometrical parameters in Figure \ref{soluzioner}.
It displays a linear variation in the homogeneously deformed core substrate and a slight nonlinear variation in the thin coating film, together with a zoom in that region for clarity.
\\

\begin{figure}[htbp]
    \centering
    \includegraphics[width=0.6\textwidth]{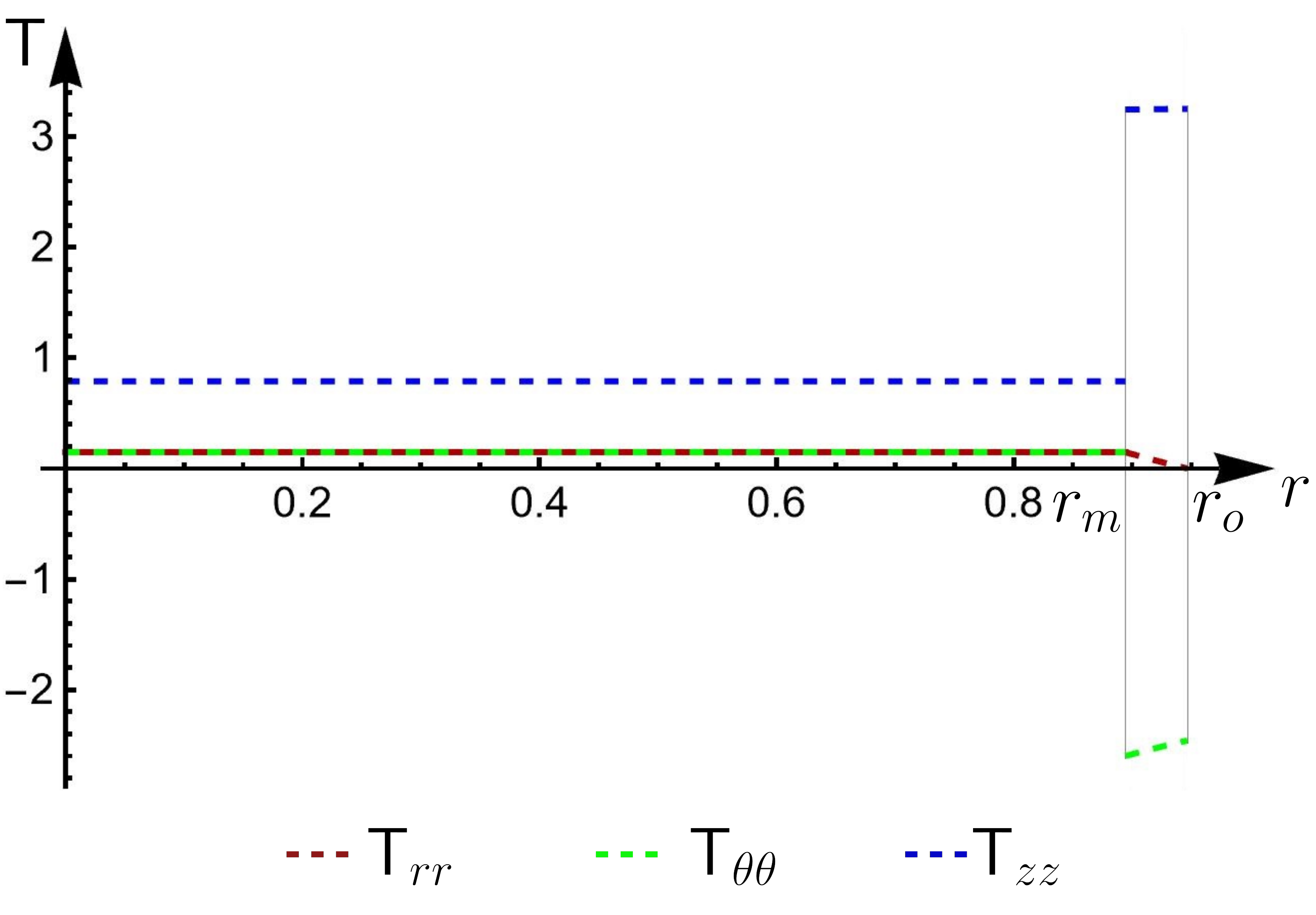}
    \caption{Variations of the Cauchy stress components with the current radius $r$. All stress components are positive (tensile) and constant in the homogeneously stretched substrate.  The radial stress (in red) is continuous across the two layers and positive (tensile) in the film, except at the external surface, where is zero.
    The axial stress (in blue) is not continuous and is positive (tensile) in the film. The circumferential stress (in green) is not continuous across the layers and is negative (compressive) in the film. The following representative parameters were chosen for this example: $\lambda=1.25$, $R_m=1$, $R_o=1.05$, $\nu_f=-0.5$, $\mu_f/\mu_s=10$, $\alpha_s=0.2$, $\alpha_f=0.5$.}
    \label{soluzionestress}
\end{figure}

%\begin{figure}[htbp]
 %   \centering
  %  \includegraphics[width=0.8\linewidth]{states.png}
   % \caption{(left) The reference configuration of the undeformed bilayer.  (centre) The virtual stress-free configuration of the two layers when deformed as if they were not attached. Being incompressible, the substrate always shrinks; if the film is auxetic, it expands, while it shrinks if $\nu_f>0$. 
    %The film needs to be compressed to remain attached to the substrate. (right) Top view of the bilayer in the final configuration, where surface wrinkles occur. }
    %\label{fig:residual}
%\end{figure}
We then plot the variations of the three principal Cauchy stress components with $r$ in Figure \ref{soluzionestress}. 
The radial stress $\tens T_{rr}$ is continuous at the interface, but the circumferential and axial stresses $\tens T_{\theta\theta}$, $\tens T_{zz}$ are not. 
In particular, the axial stress is tensile in the substrate and in the film, indicating that wrinkles cannot form in the axial direction. 
On the other hand, the hoop stress $\tens{T}_{\theta\theta}$ is compressive in the film, pointing to the possible emergence of circumferential wrinkles there.

\subsection{Incremental equations}
\label{5.4}

We now use the theory of incremental deformations \cite{ogden1997non} to investigate the onset of instability, and put the incremental equations of equilibrium in the Stroh form \cite{destrade2009bending, destrade2010deforming}, as  first-order ordinary differential equations for the incremental mechanical displacement $\mathbf u = \Re\{\mathbf U(r) \text e^{\text i m \theta}\}$ and incremental nominal traction $\mathbf s = \Re\{\mathbf S(r) \text e^{\text i m \theta}\}$, where $m\ge 2$ is the (integer) circumferential wavenumber. 
It reads 
\begin{equation}
\frac{\text d\boldsymbol{\eta}^i}{\text dr}=\frac{1}{r}N^i\boldsymbol{\eta}^i \qquad (i=s,f),
\end{equation}
where $\boldsymbol{\eta}=(\mathbf U, r \mathbf S)^\text{T}$ is the Stroh vector and $N$ is the Stroh matrix. 
%We present the explicit expressions of $N^s$ and $N^f$ in the appendix.

The components of $N^s$ are constant \cite{balbi2020mechanics}, allowing us to compute the corresponding eigenvalues and eigenvectors analytically. 
Keeping only the positive eigenvalues (so that the solution does not blow up at $r=0$) we find the solution in the form
\begin{equation}
\boldsymbol{\eta}^s (r) = c_1 r^{m-1} \boldsymbol{\eta}^s_1 + c_2 r^{m} \boldsymbol{\eta}^s_2 + c_3 r^{m+1} \boldsymbol{\eta}^s_3,    
\end{equation}
where $c_1$, $c_2$, $c_3$ are constants, and $\boldsymbol{\eta}^s_1$, $\boldsymbol{\eta}^s_2$, $\boldsymbol{\eta}^s_3$ are the eigenvectors associated to the eigenvalues $m-1$, $m$ and $m+1$.
% The eigenvalues are 
% $$ \lambda_1=1-m,\,\,\lambda_2=m-1,\,\,\lambda_3=-m,\,\,\lambda_4=m,\,\,\lambda_5=-1-m,\,\,\lambda_6=m+1$$
% while the eigenvectors that we will use in the following are
% $\boldsymbol{\eta}^\text s_2$, $\boldsymbol{\eta}^\text s_4$ and $\boldsymbol{\eta}^\text s_6$.\\

The expression for $N^f$ is not reproduced as it is too lengthy, because the film is compressible and because the components of $\tens{F}^f$ are not known explicitly but calculated from computing $r^f(R)$ numerically. 
% Analogously to what was done before, the Stroh matrix $N^c$ for the coating is again a $6\times 6$ matrix such that 
% $$
% \frac{d\boldsymbol{\eta}^c}{dr}=\frac{1}{r}N^c\boldsymbol{\eta}^c
% $$
% where $\boldsymbol{\eta}^c=(U^c(r),V^c(r),W^c(R),rp^c_{rr}(r),rp^c_{r\theta}(r),rp^c_{rz}(r))^T$. The matrix $N^c$ depends on $r$.\\
%The complete expressions of the Stroh matrices $N^\text s$ and $N^c$ and the eigenvectors are reported in Appendix A.4.\\
% We remark that both Stroh matrices respect the known symmetries, i.e. $N_4^i=-{N_1^i}^T$ and $N^i_3$ is symmetric, where $N_1$, $N_2$, $N_3$ and $N_4$ are the four blocks of the matrices.\\

Finally, the boundary conditions of continuous displacement and traction at the substrate-film interface, and traction-free outer surface, read
\begin{equation}
    \boldsymbol{\eta}^s(r_m) = \boldsymbol{\eta}^f(r_m), \qquad \mathbf S(r_o) = \mathbf 0.
\end{equation}

As is known from elastic wave theory, we may reformulate the boundary value problem using $Z^s$ and $Z^f$, the surface impedance matrices for the substrate and film, respectively. 

In the substrate, the surface impedance matrix is constant \cite{balbi2020mechanics} and we have the explicit expression $Z^s = \text i BA^{-1}$ where $A$ and $B$ are the $3 \times 3$ top and bottom submatrices of the $6 \times 3$ matrix of eigenvectors $[\boldsymbol{\eta}^s_1 |\boldsymbol{\eta}^s_2 | \boldsymbol{\eta}^s_3]$.

In the film, we compute $Z^f = Z^f(r)$ as the stable solution to the following differential Riccati equation
\begin{equation}
\dfrac{\text dZ^f}{\text dr}=\frac{1}{r}\left( N_3^f + N_4^f Z^f - Z^f N_1^f -Z^f N_2^f Z^f\right),
\end{equation}
where $N_j$ denote the classical $3\times 3$ blocks of the Stroh matrix, which we integrate numerically (having interpolated the components of $\tens F^f$) from the condition of continuity at the interface, $Z^f(r_m)=Z^s$, to the condition of zero traction at the outer face, $\det Z^f(r_o)=0$.

% We use this last relation as initial condition for the integration of the Riccati equation in the coating from $r_m$ to $r_o$ until the convergence criterion 
% $
% \det Z^c(r_o,r_o)=0
% $
% is satisfied. %From equation \eqref{basta}, it is clear that this last condition corresponds to the traction-free boundary condition at the exterior radius of the coating.\\

To run the numerical code, we fix the values of $H$, $\mu$, $\nu$, $\alpha_s$ and $\alpha_f$ and we generate two iterative loops over $m$ and $\lambda$. 
Inside the two loops we integrate the Riccati equation with the initial condition on the interface, until the stop condition at $r_o$ is satisfied. 
This algorithm yields the critical thresholds $\lambda_\text{cr}$ and associated wavenumbers $m_\text{cr}$ corresponding to the minimum stretch ratio determined by the stop condition for all $m$.

\subsection{Finite Element models}
\label{5.5}

%%%%%%%%%%%%%%%%%%%%%%%%%%%%%%%%%%%%%%

We construct two sets of finite element (FE) models of the coated
cylinders in ABAQUS. The first set is used to validate the
semi-analytical stability analysis and represents the coating as a
homogeneous compressible material. The second set explicitly resolves
an architectured periodic coating containing elliptical voids,
designed to produce an effective auxetic response
(Section~\ref{sec:architectured}). The substrate radius is normalised
to $R_m=1$ in both cases.

The homogeneous bilayered cylinder models are constructed and meshed
directly in ABAQUS/CAE. The compressible film is discretised using
20-node quadratic reduced-integration hexahedral elements
(\textsf{C3D20R}), whereas the nearly incompressible substrate is
discretised using 20-node quadratic reduced-integration hybrid
hexahedral elements (\textsf{C3D20RH}). The final homogeneous mesh
contains $128\,602$ nodes and $30\,192$ elements, comprising
$5\,328$ \textsf{C3D20R} film elements and $24\,864$
\textsf{C3D20RH} substrate elements.

\begin{figure}
    \centering
    \includegraphics[width=0.7\linewidth]{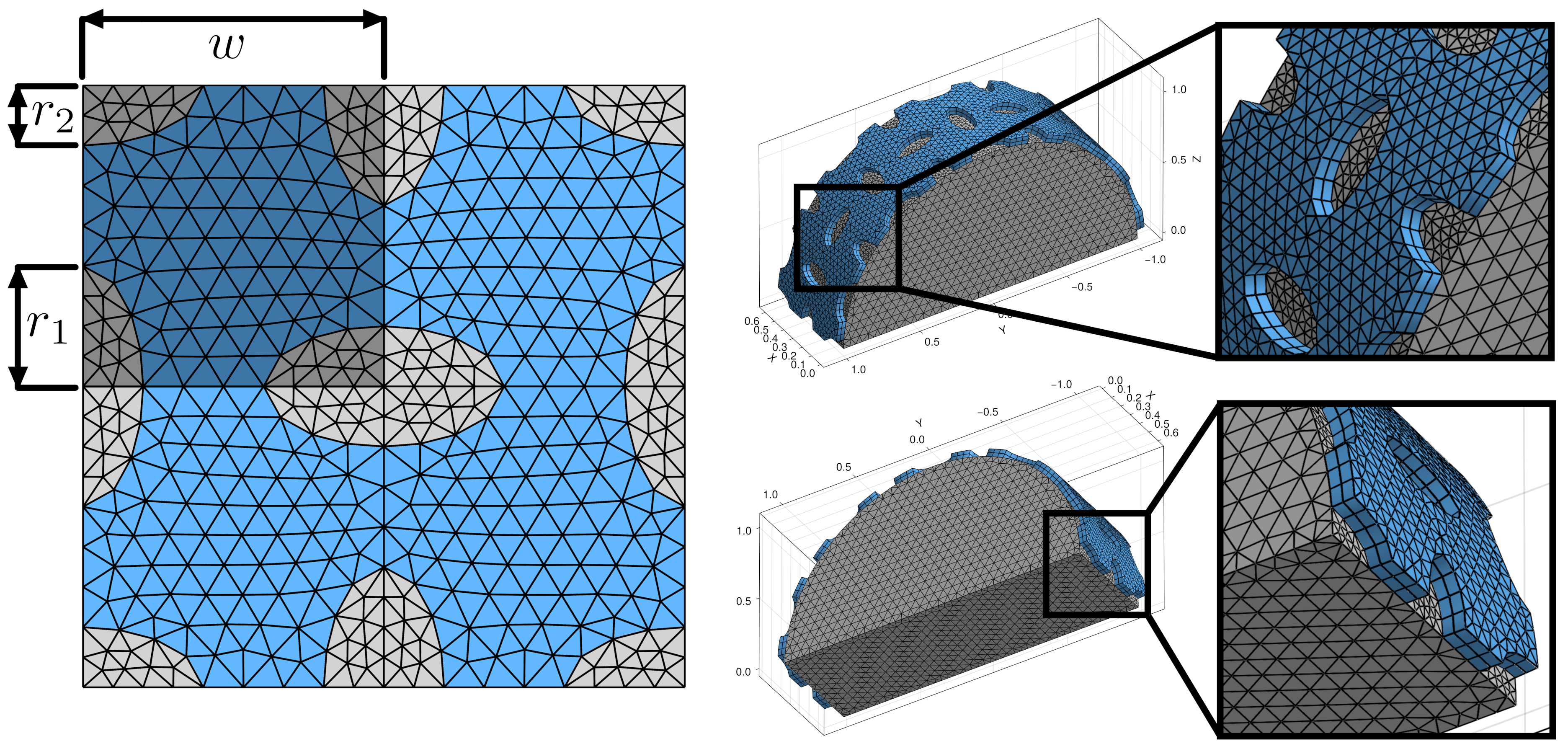}
    \caption{Generation of the architectured finite element model.
    (left) Two-dimensional periodic mesh containing elliptical
    features with major radius $r_1$ and minor radius $r_2$.
    (right) Final three-dimensional mesh comprising the
    \textsf{C3D15} architectured coating and the
    \textsf{C3D10H} substrate. The inset views illustrate the
    formation of the coating by thickening the triangulated outer
    surface.}
    \label{fig:fea_mesh}
\end{figure}

In contrast, the geometry and mesh of the architectured model are
generated parametrically using a custom pipeline written in Julia
\cite{bezanson2017julia}, built on the computational mechanics
library Comodo.jl \cite{moerman_comodojl_2025}, ABAQUSTools.jl
\cite{moerman_abaqustoolsjl_2026}, and the visualisation library
Makie.jl \cite{danisch_makiejl_2021}. The pipeline constructs the
periodic microstructure, generates the coating and substrate meshes,
identifies the node and element sets required for the boundary
conditions, and writes the corresponding ABAQUS input file. The full
mesh-generation procedure is described in
Appendix~\ref{app:fe_pipeline}. The geometrical and mesh parameters
used for the architectured model are listed in
Table~\ref{table:fea_parameters}, and the resulting mesh is shown in
Figure~\ref{fig:fea_mesh}.

\begin{table}[htbp]
\centering
\small
\renewcommand{\arraystretch}{1.25}
\setlength{\tabcolsep}{5pt}
\begin{tabular}{@{}clcc@{}}
\toprule
Parameter & Description & Value & Unit \\
\midrule
$n_1$ & Number of axial repetitions           & $1$          & $[-]$ \\
$n_2$ & Number of circumferential repetitions & $5$          & $[-]$ \\
$w$   & Unit-cell half-width                  & $\pi/(2n_2)$ & mm \\
$r_1$ & Ellipse major radius                  & $2w/5$       & mm \\
$r_2$ & Ellipse minor radius                  & $w/5$        & mm \\
$t$   & Auxetic-layer thickness               & $1/20$       & mm \\
$s_1$ & Layer node spacing                    & $2w/15$      & mm \\
$s_2$ & Cylinder boundary node spacing        & $3s_1/2$     & mm \\
$f$   & Maximum tetrahedral volume factor     & $10$         & $[-]$ \\
\bottomrule
\end{tabular}
\caption{Geometrical and mesh parameters used to generate the
architectured finite element model.}
\label{table:fea_parameters}
\end{table}

The architectured coating and substrate form a conforming mesh with
shared nodes across their common interface. The nearly incompressible
substrate is discretised using 10-node quadratic hybrid tetrahedral
elements (\textsf{C3D10H}), while the architectured coating is
discretised using 15-node quadratic wedge elements
(\textsf{C3D15}). The final architectured model contains
$59\,102$ nodes and $33\,256$ elements, comprising $27\,816$
\textsf{C3D10H} substrate elements and $5\,440$
\textsf{C3D15} coating elements.

For both the homogeneous and architectured models, periodic boundary
conditions are imposed between the two opposite axial end faces.
Corresponding nodes on these faces are related through multi-point
equation constraints of the form
\begin{equation}
    u_i^{+}-u_i^{-}-u_i^{\mathrm{ref}}=0,
    \qquad i=1,2,3,
    \label{eq:fe_pbc}
\end{equation}
where $u_i^{+}$ and $u_i^{-}$ denote the displacement components of
corresponding nodes on the two axial end faces, and
$u_i^{\mathrm{ref}}$ denotes the corresponding displacement component
of a reference point. These constraints prescribe the relative
displacement between the two periodic faces without rigidly coupling
the individual face nodes. The transverse components of the reference
point are constrained as
$u_2^{\mathrm{ref}}=u_3^{\mathrm{ref}}=0$, while the imposed axial
extension is prescribed through $u_1^{\mathrm{ref}}$.

\begin{figure}[hbtp]
    \centering
    \includegraphics[width=0.6\textwidth]{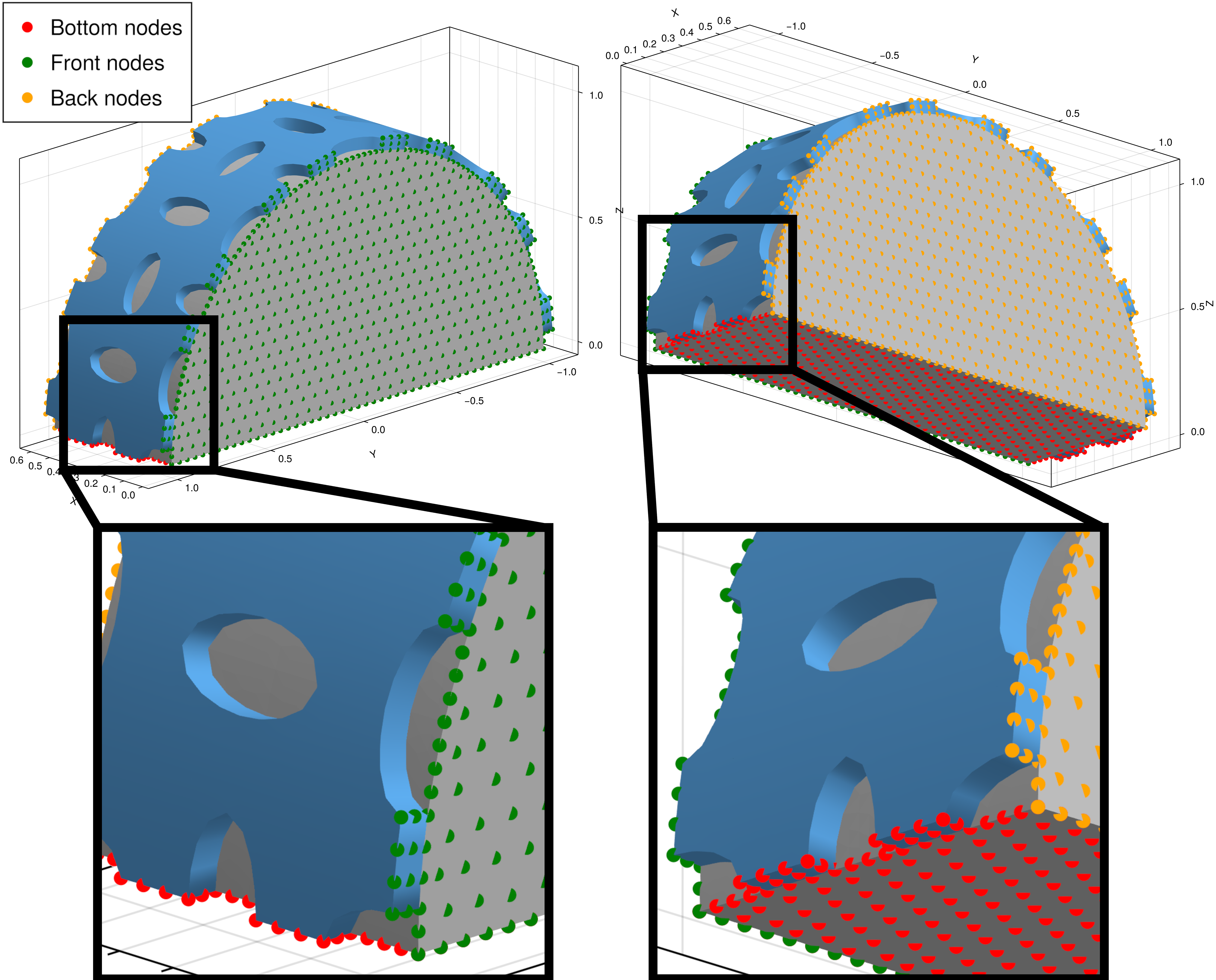}
    \caption{Node sets used to prescribe the boundary conditions of
    the architectured finite element model. The red nodes identify
    the symmetry plane, while the green and yellow nodes identify the
    two opposite axial end faces used to impose the periodic
    constraints.}
    \label{fig:fea_bc}
\end{figure}

A symmetry condition is imposed on the diametral symmetry plane. To
eliminate the remaining rigid-body translations, a central node on
one axial end face is additionally restrained in the $u_1$ and $u_3$
directions. Figure~\ref{fig:fea_bc} illustrates the corresponding
node sets for the architectured model.

For each material case, a geometrically nonlinear static pre-stretch
is first applied to establish the finitely deformed and prestressed
base configuration. The magnitude of the pre-stretch is selected
separately for the different film stiffnesses considered. Under axial
extension, the nearly incompressible substrate contracts laterally,
whereas the coating exhibits a different lateral response. This
mismatch generates compressive circumferential stress in the coating,
which provides the prestress driving the subsequent wrinkling
instability.

The onset of instability is then determined using the linear
perturbation \textsf{BUCKLE} procedure in ABAQUS
\cite{bilayercao}. If $d_{\mathrm{pre}}$ denotes the axial
reference-point displacement applied during the nonlinear pre-stretch,
$d_{\mathrm{b}}$ denotes the prescribed axial displacement defining
the buckling perturbation, and $\eta$ is the corresponding buckling
eigenvalue, the critical axial displacement is
\begin{equation}
    d_{\mathrm{cr}}
    =
    d_{\mathrm{pre}}+\eta d_{\mathrm{b}},
    \label{eq:fe_critical_displacement}
\end{equation}
and the associated critical stretch is
\begin{equation}
    \lambda_{\mathrm{cr}}
    =
    1+\frac{d_{\mathrm{cr}}}{L},
    \label{eq:fe_critical_stretch}
\end{equation}
where $L$ is the initial axial length of the model. This formulation
allows the pre-stretch and perturbation displacement to be selected
independently for each material case while determining the critical
stretch from the corresponding eigenvalue.
The Julia scripts used to generate the architectured finite element
model and its ABAQUS input files are available open source at
\url{https://github.com/COMODO-research/Auxetic_Bilayered_Cylinder/tree/review-matprop-pbc-fixes}.

%%%%%%%%%%%%%%%%%%%%%%%%%%%%%%%%%

\section{Results}
\label{5.6}

%%%%%%%%%%%%%%%%%%%%%%%%%%%%%%%%

\subsection{Validation of the Finite Element discretisation}
\label{sec:fevalidation}

\begin{figure}[h!]
    \centering
    \includegraphics[width=0.6\linewidth]{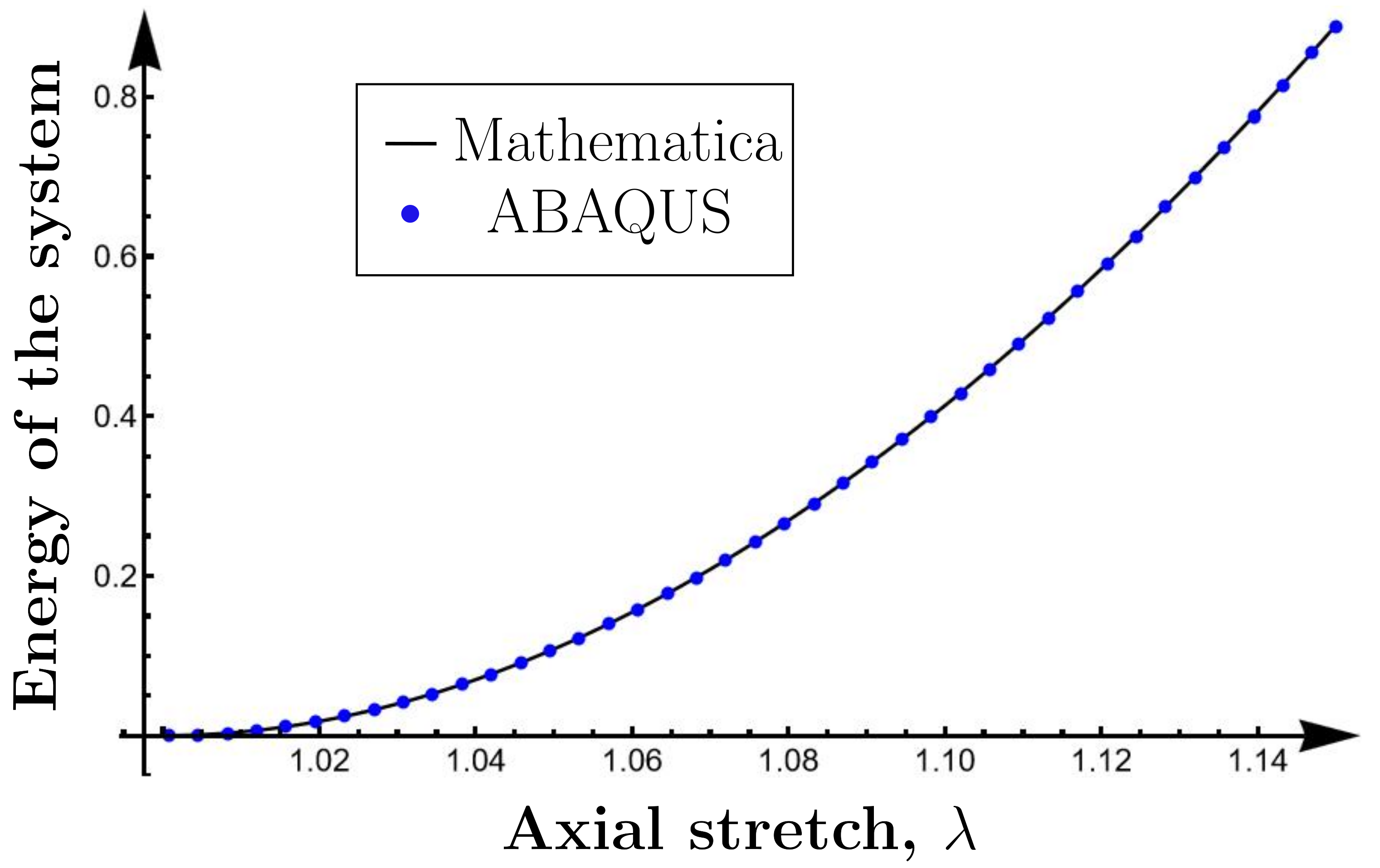}
    \caption{Comparison between the results of Mathematica and ABAQUS for the energy developed by the system when it undergoes an axial tension of stretch ratio $\lambda$ in the interval $[1,1.15]$. 
    Here the other parameters are $\mu_s=1$, $\mu_f=50$, $\alpha_s=0.2$, $\alpha_f=0.5$, $H=1.05$, $\nu=-0.7$. }
    \label{fig:energy}
\end{figure}

To validate the Finite Element discretisation of the coated cylinders, we compute the energy developed by the structure under axial tension and compare the values given by Mathematica (exact solution) and by ABAQUS (FE simulation).
Figure \ref{fig:energy} shows excellent agreement, with differences of less than 0.14\%.

%++++++++++++++++++++++++

\subsection{Onset of wrinkling: semi-analytical predictions and Finite Element validation}
\label{sec:onset}

%++++++++++++++++++++

We use a semi-analytical approach through Mathematica to find thresholds for surface instability of the exterior layer, in particular circumferential wrinkles.
For our study, we fixed the material parameters at $\alpha_f=0.5$ and $\alpha_s=0.2$, the geometrical parameter $H=R_o/R_m$ as $H=1.05$, and took $\mu=\mu_f/\mu_s = 50,100,500$ for the film-substrate shear moduli ratio. 

\begin{figure}[htbp]
\begin{subfigure}{.49\textwidth}
		\centering
		\includegraphics[width=0.99\textwidth]{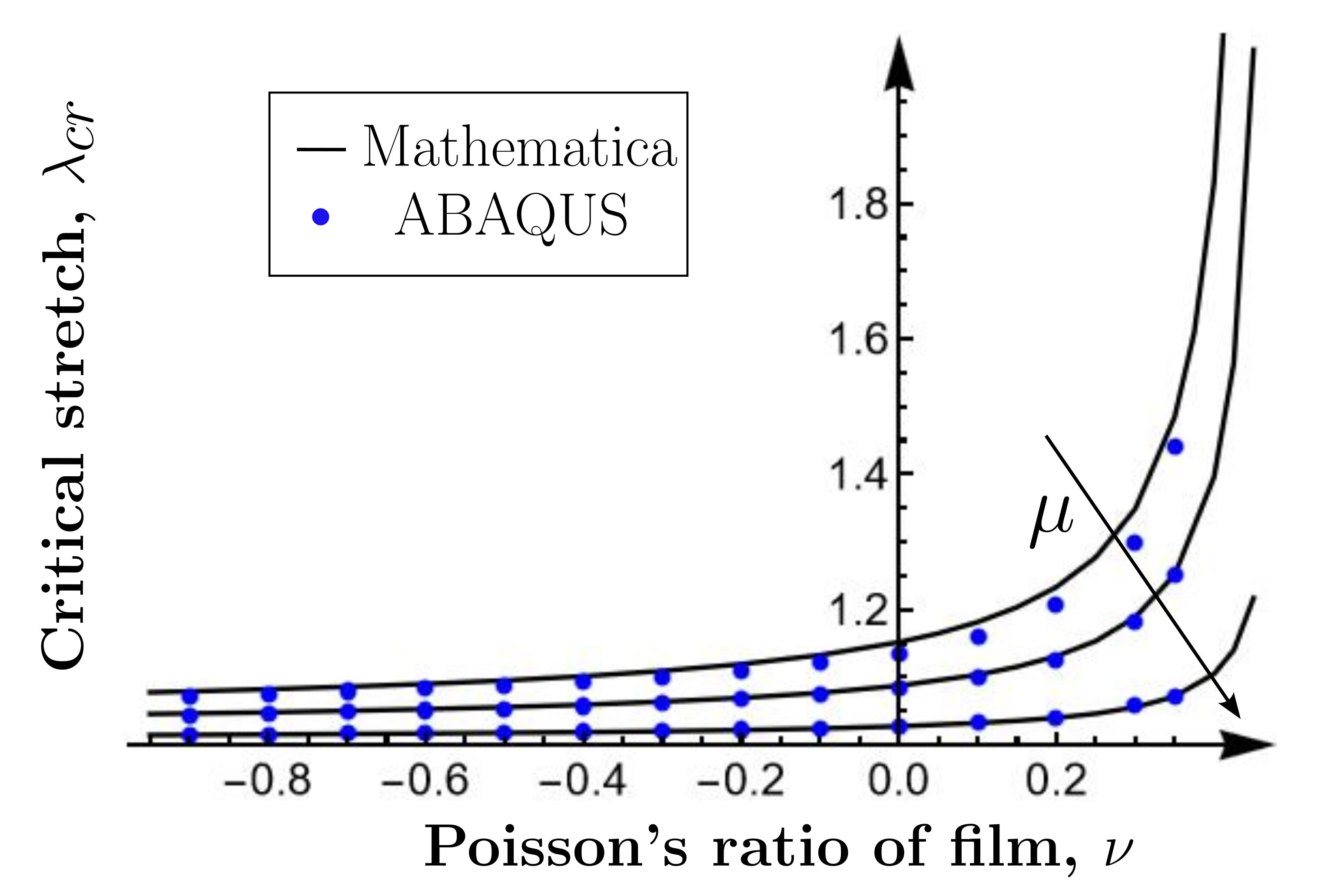}
        \caption{Plot of the critical thresholds $\lambda_{cr}$ against the Poisson ratio $\nu$ of the film. We notice that $\lambda$ increases as $\nu$ increases and decreases as $\mu$ increases.}
        \label{lambdaresults}
        \end{subfigure}
        \hfill
        \begin{subfigure}{.49\textwidth}
\centering
		\includegraphics[width=0.99\textwidth]{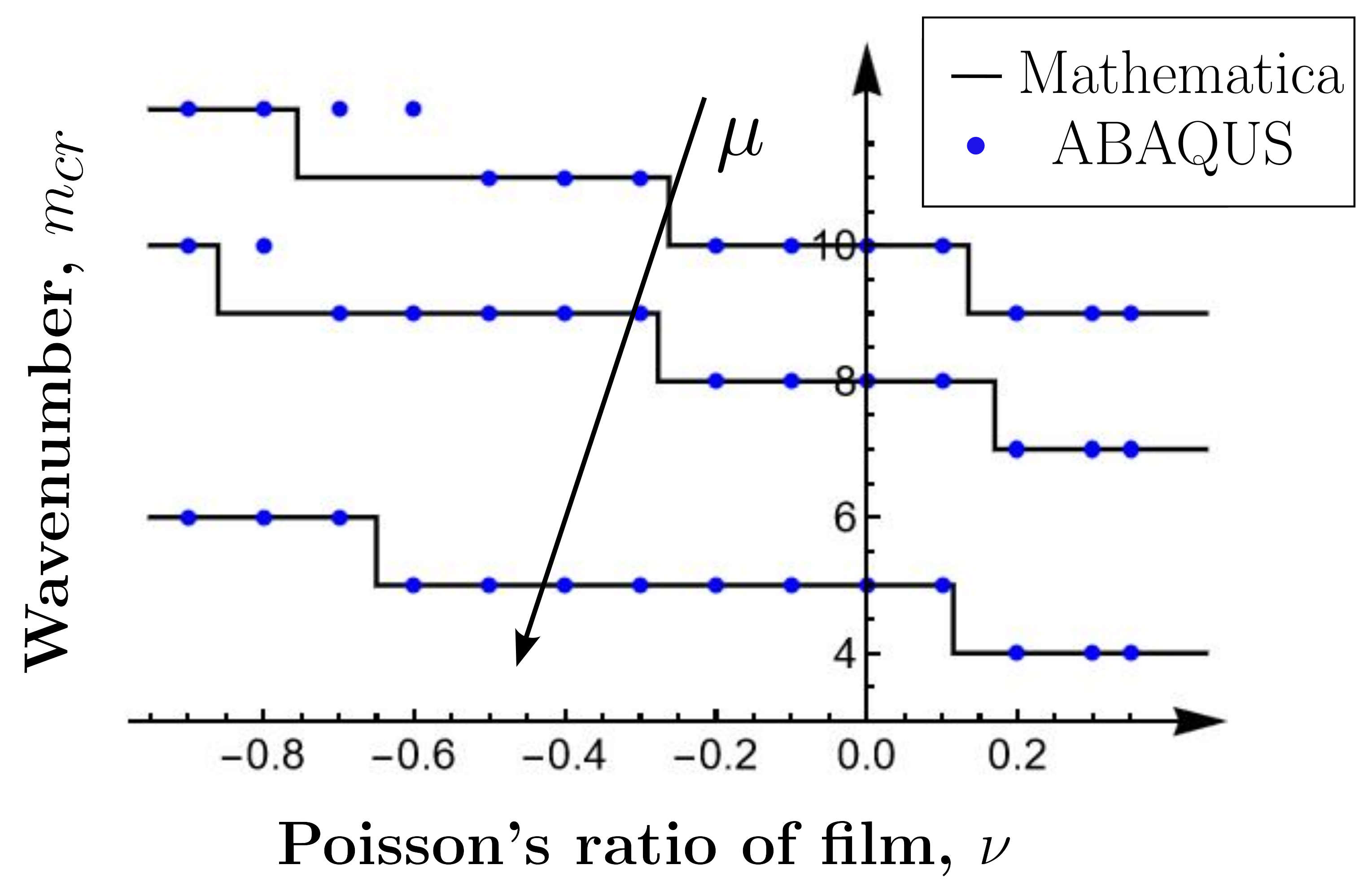}	
        \caption{Plot of the wavenumbers $m_{cr}$ associated to the critical thresholds $\lambda_{cr}$ against the Poisson ratio $\nu$ of the film. We notice that $m$ decreases as $\nu$ increases and as $\mu$ increases. }
        \label{mresults}
        \end{subfigure}
		\caption{Here the Poisson ratio $\nu$ of the film varies from $-0.95$ to $0.45$, while $\alpha_f=0.5$, $\alpha_s=0.2$, $H=1.05$, $\mu =50, 100, 500$. The blue dots represent the results of ABAQUS simulations for the same material parameters.}
	\end{figure}

Figure \ref{lambdaresults} shows the critical thresholds $\lambda_\text{cr}$ against the Poisson ratio $\nu$ of the coating and Figure \ref{mresults} shows the variations of the corresponding wavenumbers $m_\text{cr}$.
We see that $\lambda_\text{cr}$ increases with $\nu$, with rapid blow up as the Poisson ratio of the coating tends to that of the incompressible substrate in the $\nu\to 0.5$ limit.
Also, we note that increasing the shear ratio contrast $\mu$ leads to lower critical stretch ratios.
Figure \ref{mresults} highlights that increasing the value of $\nu$ decreases the critical circumferential wavenumber $m_\text{cr}$ and so, increases the wavelength, and similarly when increasing $\mu$. 
We also checked that when $\mu<1$ (substrate stiffer than coating), there are no instabilities developing on the exterior surface.

%As said above, as $\nu$ tend to $0.5$ the two materials tends to not develop singularities and then to a lower wavenumber.

%: just as an example, for the blue curve obtained for $\mu=50$, $\lambda$ ranges between $1.077$ for $\nu=-0.95$, i.e. for a highly auxetic coating, and $3.73$ for $\nu=0.45$. We remark that the plotrange of the graph is the interval $[1,2.12]$ just to enhance the legibility of the four different curves, where by example the $\lambda_{cr}$ of the red curve for $\mu=1000$ assumes values in $[1.009.125]$.\\

\begin{figure}[h!]
        \centering\includegraphics[width=0.9\textwidth]{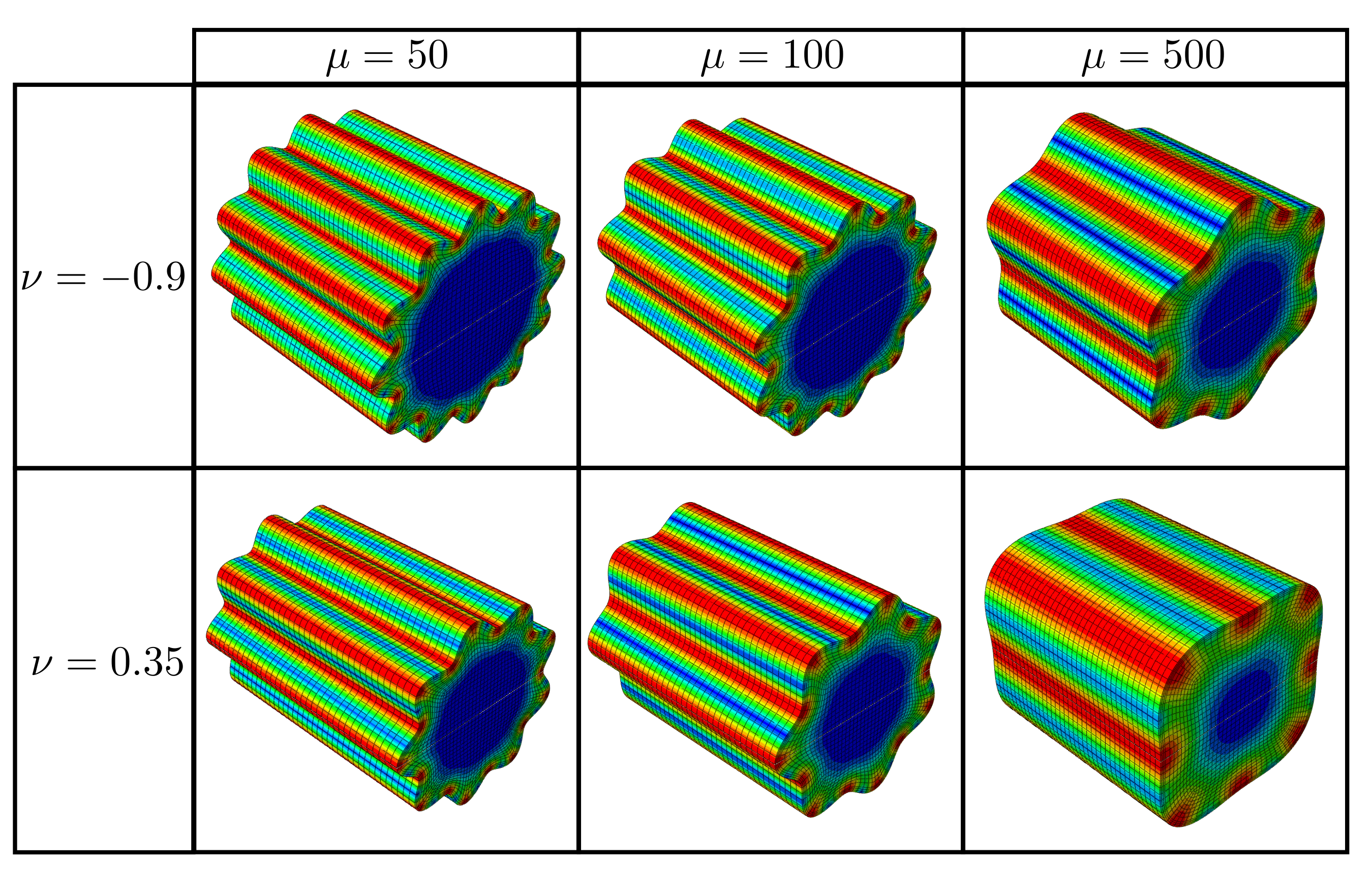}
    \caption{Finite Element simulations for the wrinkled configurations of an incompressible cylinder coated with a homogenous compressible film under tension, for different  values of the  shear modulus ratio $\mu = \mu_f/\mu_s$ and Poisson ratio $\nu = \nu_f$.}
    \label{ABAQUS}
\end{figure}

We then validate the semi-analytical results against finite element simulations. 
When the film is uniform and made of a compressible Blatz-Ko material, we obtain the blue dots in Figure \ref{lambdaresults} and \ref{mresults} for the critical stretches and associated wavenumbers.
The agreement between the two approaches is strong, with less than 2\% difference. 
Figure \ref{ABAQUS} displays some buckled configurations obtained in ABAQUS for different $\nu$ and $\mu$.

These trends are in good qualitative agreement with those obtained by Venkata et al.~\cite{venkata2025wrinkling} for the analogous rectangular bilayer under uniaxial tension. 
For the near-incompressible substrate limit considered in Figures~4(a)--(b) of \cite{venkata2025wrinkling}, the critical stretch and wavenumber decrease with increasing $\nu$ in the same way as found here, and Figure~6 of \cite{venkata2025wrinkling} shows the same decrease in both $\lambda_\text{cr}$ and $m_\text{cr}$ as the modulus ratio increases from $\mu=50$ to $\mu=500$, mirroring our results in Figures~\ref{lambdaresults} and \ref{mresults}. The main qualitative difference introduced by the cylindrical geometry is that the circumferential wavenumber must be a positive integer, so that, unlike in the planar case, the marginal stability condition is only sampled at discrete values of $m$; we return to this point in the Conclusions.

%+++++++++++++++++++++++++++++++++

% \subsection{Finite Element simulations of architectured auxetic coatings}
% \label{sec:architectured}

% %++++++++++++++++++++++++++++

% \begin{figure}[htbp]
%     \centering
%     \includegraphics[width=0.48\textwidth]{FrontView-Aux-ShM50-Buckling.png}
%     \hfill
%     \includegraphics[width=0.48\textwidth]{Isometric-Aux-ShM50-Buckling2.png}
%     \caption{Caption figure 9}
%     \label{fig:auxetic1}
% \end{figure}
% \begin{figure}[htbp]
%     \centering
%     \includegraphics[width=0.48\textwidth]{uRadial-Disp-Aux-ShM50-Prestretch.png}
%     \hfill
%     \includegraphics[width=0.48\textwidth]{uZ-Disp-Aux-ShM50-Prestretch.png}
%     \caption{Caption}
%     \label{fig:auxeticdisp}
% \end{figure}

% \color{blue}[architectured-film simulations]\color{black}

% The finite-element simulations therefore play two roles. First, the homogeneous-film simulations provide a direct validation of the incremental stability calculation, both at the level of the finite deformation and at the level of the critical stretch and circumferential wavenumber. Second, the architectured-film simulations show that a geometrically realizable auxetic coating can reproduce the qualitative instability mechanism predicted by the continuum model. The latter simulations also guide the design of the 3D-printed specimens used in the experimental trials.

%++++++++++++++++++++++++++++++++
\subsection{Finite Element simulations of architectured auxetic coatings}
\label{sec:architectured}

%++++++++++++++++++++++++++++++++

Having validated the homogeneous film model against the
semi-analytical predictions in the preceding section, we now consider
an explicitly architectured auxetic coating. The periodic
microstructure described in Section~\ref{5.5} has an effective Poisson
ratio of approximately $\nu_{\mathrm{eff}}=-0.02$. 

For a film-to-substrate shear modulus ratio $\mu=\mu_f/\mu_s=50$, 
the architectured finite element model predicts a critical stretch $\lambda_{\mathrm{cr}}^{\mathrm{arch}} \simeq 1.249$,
with corresponding circumferential wavenumber $m_{\mathrm{cr}}^{\mathrm{arch}}=10$.

For comparison, the critical stretch of the homogeneous-film model at
$\nu_f=-0.02$ is estimated by linear interpolation between the finite
element results obtained at $\nu_f=-0.1$ and $\nu_f=0$. Using $\lambda_{\mathrm{cr}}(-0.1)=1.1202$, $\lambda_{\mathrm{cr}}(0)=1.1355$, gives
$\lambda_{\mathrm{cr}}^{\mathrm{hom}}(-0.02)
    \simeq 1.132$.
The homogeneous simulations at both neighbouring values of $\nu_f$
give $m_{\mathrm{cr}}=10$, in agreement with the mode obtained for the
architectured coating. The interpolated homogeneous critical stretch
is approximately $9.4\%$ lower than the architectured value when
normalised by $\lambda_{\mathrm{cr}}^{\mathrm{arch}}$.

\begin{figure}[htbp]
    \centering
    \includegraphics[width=0.48\textwidth]{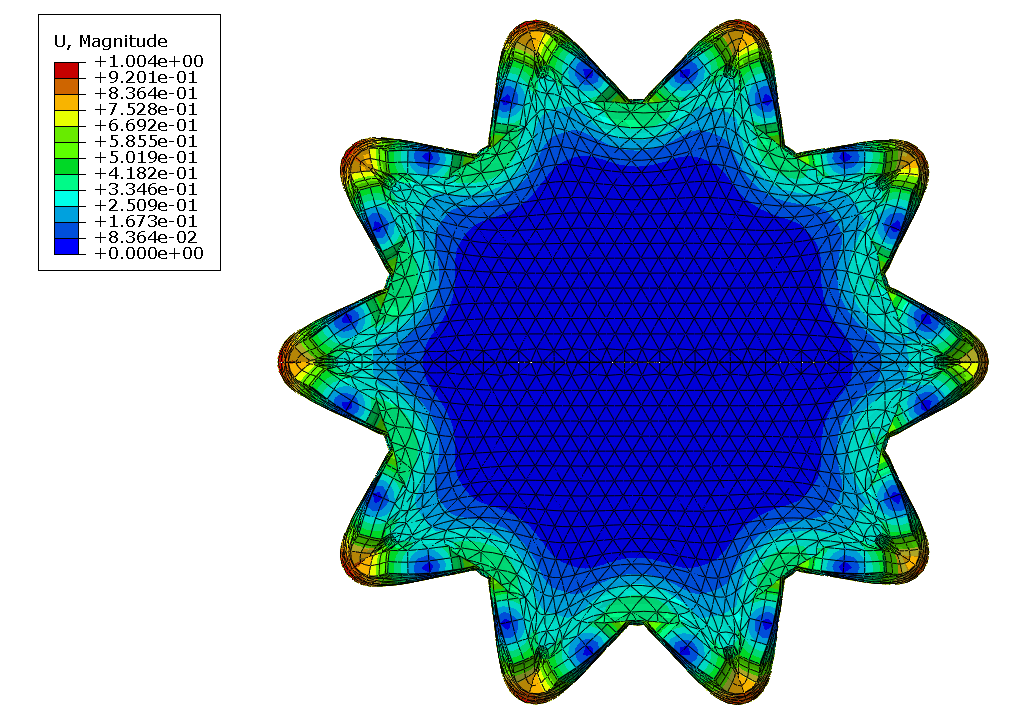}
    \hfill
    \includegraphics[width=0.48\textwidth]{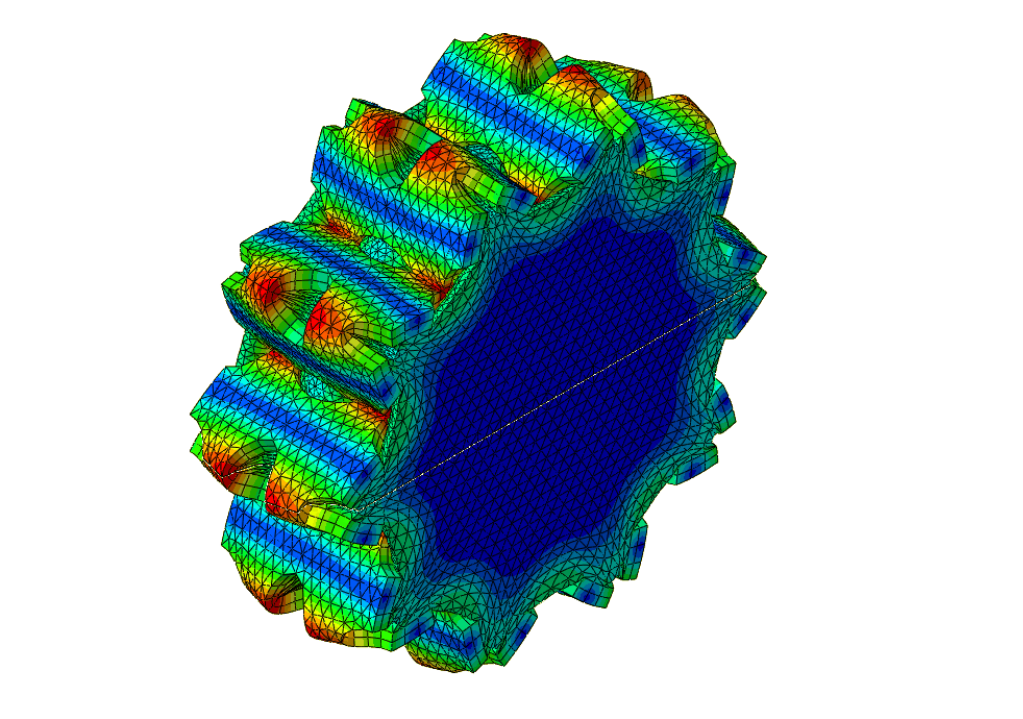}
    \caption{Critical buckling mode of the cylinder with an
    architectured auxetic coating for $\mu_f/\mu_s=50$, shown in
    front and isometric views. The contours show the magnitude
    $U=\lVert\mathbf{u}\rVert$ of the normalised displacement
    eigenvector obtained from the linear perturbation buckling
    analysis. The eigenvector amplitude is arbitrary and represents
    the spatial form of the instability rather than a physical
    post-buckling displacement. The corresponding circumferential
    wavenumber is $m_{\mathrm{cr}}=10$.}
    \label{fig:auxetic1}
\end{figure}

Figure~\ref{fig:auxetic1} shows the critical buckling mode of the
architectured model. The mode consists of ten circumferential
undulations, corresponding to $m_{\mathrm{cr}}=10$. The contours show
the magnitude of the displacement eigenvector obtained from the linear
perturbation buckling analysis. Since the eigenvector is normalised,
its amplitude is not a physical post-buckling displacement, but
describes the spatial form of the instability.

\begin{figure}[htbp]
    \centering
    \includegraphics[width=0.48\textwidth]{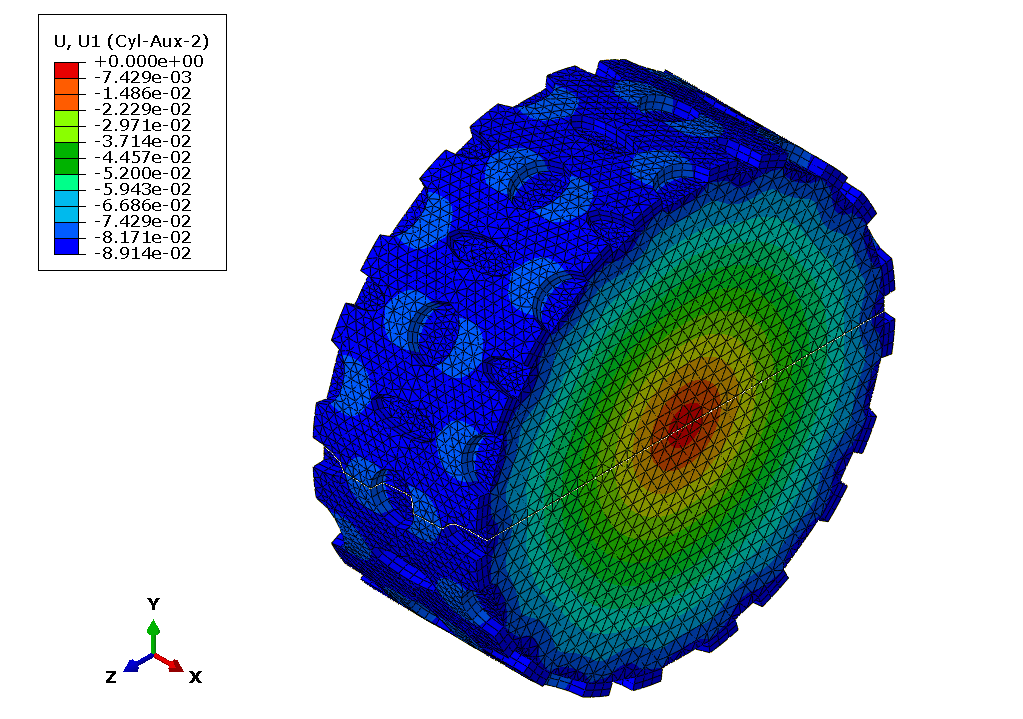}
    \hfill
    \includegraphics[width=0.48\textwidth]{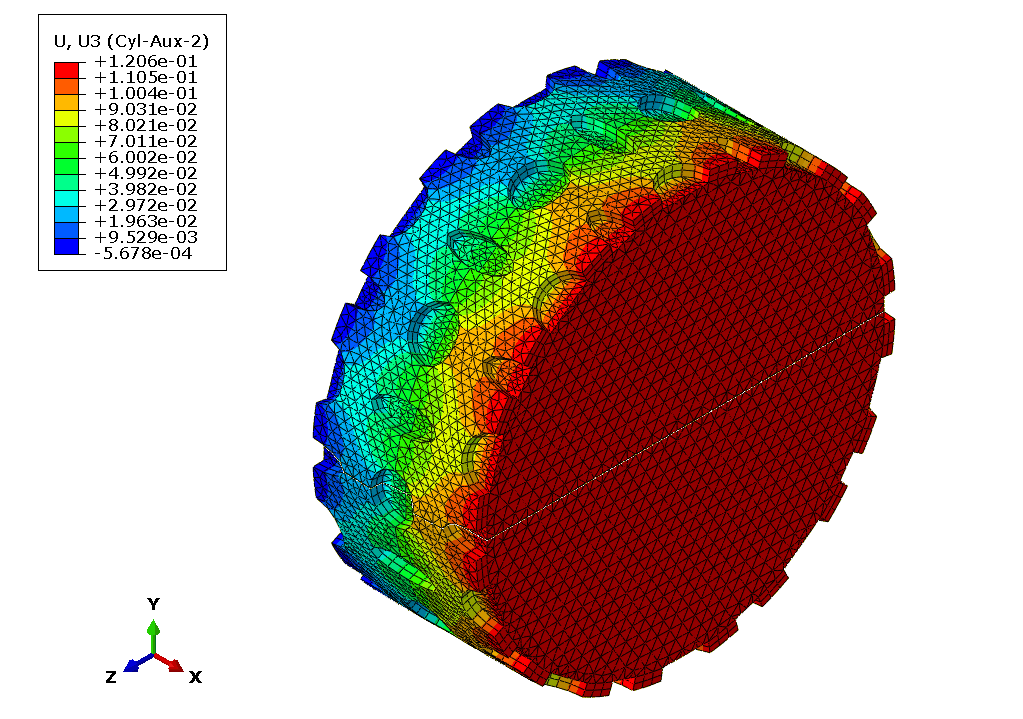}
    \caption{Displacement field of the architectured bilayered
    cylinder at the end of the prestretch step for
    $\mu_f/\mu_s=50$: (left) radial displacement $u_r$
    (\textsf{U1}) and (right) axial displacement $u_z$
    (\textsf{U3}), expressed in the cylindrical coordinate system.
    This configuration provides the finitely deformed and prestressed
    base state for the subsequent linear perturbation buckling
    analysis.}
    \label{fig:auxeticdisp}
\end{figure}

Figure~\ref{fig:auxeticdisp} shows the displacement field at the end
of the nonlinear prestretch step. The radial displacement $u_r$ and
axial displacement $u_z$ are reported in the cylindrical coordinate
system and describe the finitely deformed base state used in the
subsequent buckling analysis.

The finite element simulations thus serve two purposes. The
homogeneous-film simulations provide a direct comparison with the
incremental stability analysis in terms of both the critical stretch
and the circumferential wavenumber. The architectured-film simulations
show that the same circumferential instability mode can also be
obtained when the auxetic response is generated explicitly by a
periodic microstructure. For the case considered here, the
architectured model gives the same critical wavenumber as the
neighbouring homogeneous-film simulations, while its critical stretch
is approximately $9.4\%$ higher than the interpolated homogeneous
value. The architectured model also provides the basis for the
3D-printed specimens considered in the proof-of-concept experiments
below.

%++++++++++++++++++++++++++++++++

\subsection{Preliminary 3D-printed experiments: a proof-of-concept study}
\label{sec:printed}

%+++++++++++++++++++++++++++++++

We complement the theoretical and computational results above with a proof-of-concept fabrication study, rather than an experimental validation. 
The aim here is to demonstrate that soft cylindrical specimens with stiff, architectured auxetic coatings can be 3D-printed and mechanically tested under axial tension, not to quantitatively confirm the predicted critical stretch and wavenumber.

\begin{figure}[h!]
    \centering
    \includegraphics[width=0.49\textwidth]{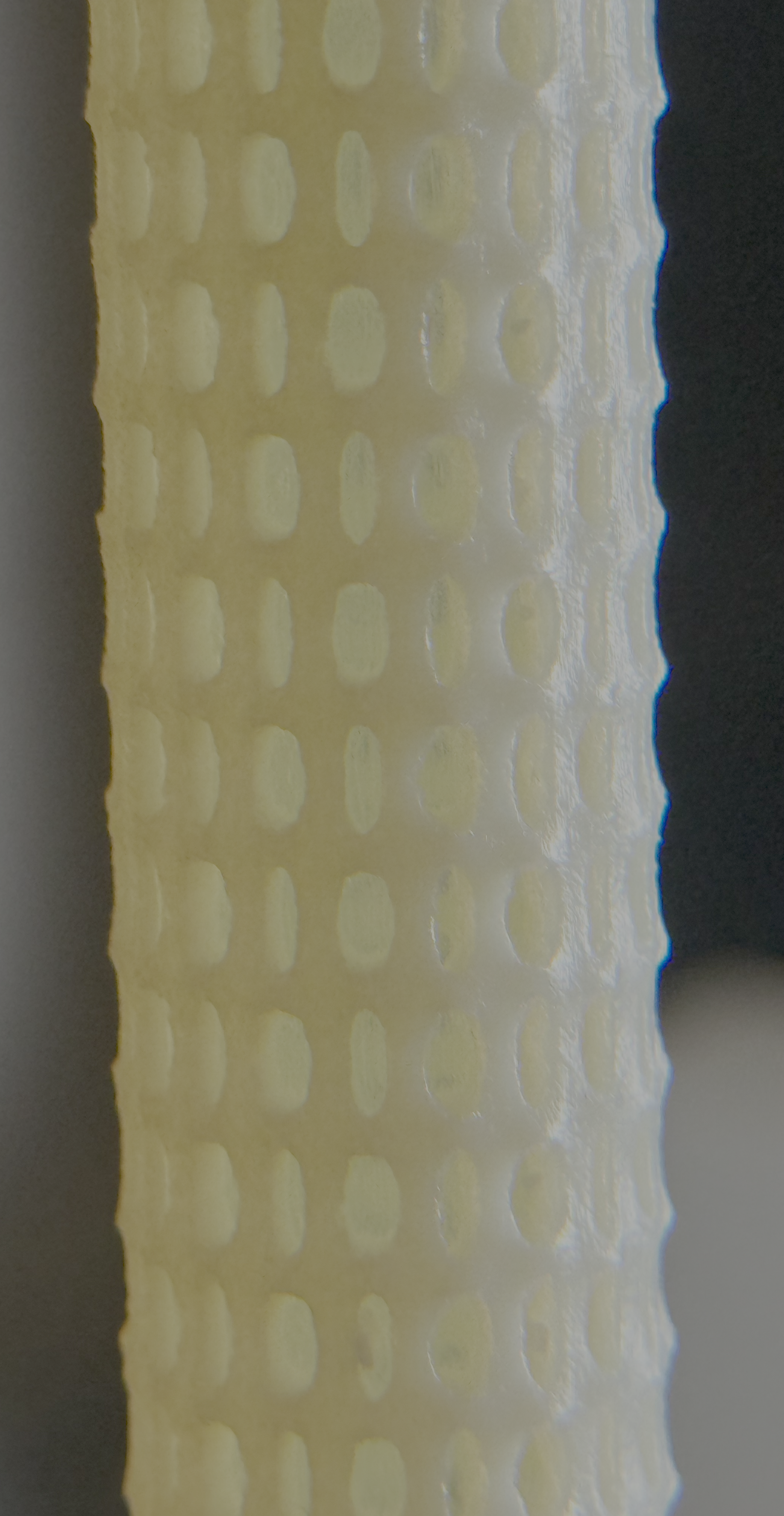}
    \caption{Preliminary 3D-printed bilayered cylinder with an architectured auxetic coating under axial extension. The regular surface pattern reflects the printed microstructure of the coating. Small surface undulations are visible near the outer contour, but their amplitude and wavelength are difficult to distinguish from the microstructural length scale. The observation is suggestive but not conclusive evidence of instability-induced wrinkling.}
    \label{fig:printed-cylinder}
\end{figure}

We fabricated bilayered cylinders in which the auxetic coating was realised as an architectured surface microstructure of the kind described in Section~\ref{sec:architectured}, and subjected them to axial extension. Under tension, the outer layer displayed small surface undulations and local changes in texture, as shown in Figure~\ref{fig:printed-cylinder}. However, the amplitude of these features was close to the resolution limit of our optical observations, and their characteristic length scale was comparable to that of the printed microstructure itself. It follows that it was not possible to determine unambiguously whether the observed surface modulation reflects a genuine tension-induced wrinkling instability, or simply the geometry and deformation of the microstructured coating.

We therefore report this fabrication study as preliminary and not conclusive.
It establishes that the specimens can be manufactured and mechanically tested, and it identifies the main obstacle to a future quantitative validation, namely that the predicted wrinkle amplitude is small relative to the printable feature size. Closing this gap will require finer printing resolution or larger specimens, together with surface metrology capable of separating the wrinkle wavelength from the unit-cell length scale of the coating.

Details of the fabrication and mechanical characterisation of the 3D-printed specimens, including the tensile stress--strain response of the coated cylinders, are given in Appendix~\ref{app:experiments}.

%%%%%%%%%%%%%
\section{Conclusions}
\label{sec:conclusions}

%%%%%%%%%%%%%%

We have investigated the onset of circumferential wrinkling in a bilayered cylinder composed of an incompressible substrate coated with a stiff, compressible film, when the structure is subjected to axial tension. Although tension is usually associated with stabilisation rather than instability, we have shown that the incompressible substrate's lateral contraction under stretch forces a compressive hoop stress onto the compressible coating, and that this mismatch drives the onset of circumferential wrinkles on the outer surface, while the axial stress remains tensile throughout both layers.

Using the theory of incremental deformations, cast in the Stroh form and integrated numerically via the differential Riccati equation for the surface impedance matrix, we obtained the critical stretch ratio $\lambda_\text{cr}$ and the associated circumferential wavenumber $m_\text{cr}$ as functions of the coating's Poisson ratio $\nu$ and the film-to-substrate shear modulus ratio $\mu$. We find that $\lambda_\text{cr}$ decreases as $\nu$ decreases toward auxetic values and as $\mu$ increases, while the associated wavenumber $m_\text{cr}$, and hence the wrinkle wavelength, responds in the opposite sense.
Also, no instability develops when the coating is softer than the substrate ($\mu<1$). These predictions are validated by Finite Element simulations of a homogeneous compressible coating, with agreement to within 0.14\% on the strain energy and 2\% on $\lambda_\text{cr}$ and $m_\text{cr}$.

These trends mirror those obtained by Venkata et al. \cite{venkata2025wrinkling} for the analogous rectangular bilayer, confirming that the same Poisson mismatch mechanism governs both geometries in the thin-coating limit $H\to1$, where curvature effects are weak. Curvature nonetheless introduces one new qualitative feature, because the wavenumber $m$ is necessarily a positive integer, so the marginal stability condition is sampled only at discrete values of $m$, rather than continuously as in the rectangular case. 
The consequences of this curvature-induced mode quantisation for mode competition and imperfection sensitivity would be worth investigating further.

Motivated by the theoretical results, we explored whether the auxetic behaviour required to promote early wrinkling can be realised with a geometrically architectured coating rather than an idealised homogeneous compressible material, and we fabricated 3D-printed bilayered cylinders with stiff, architectured auxetic outer layers. This was a proof-of-concept fabrication exercise rather than a quantitative validation. 
It confirmed that such specimens can be manufactured and mechanically tested under axial tension, while showing that the predicted wrinkle amplitude is presently too small, relative to the printed microstructure's own length scale, to be unambiguously resolved. Closing this gap through finer printing resolution, larger specimens, or higher-resolution surface metrology, is a natural next step, as is extending the present analysis to other loading modes, finite-length effects, and growth- or swelling-induced analogues of the tension case studied here.

%%%%%%%%%%%%%%%%%%%%%%
\section*{Acknowledgments}
Chiara Lonati was supported by the MICS (Made in Italy – Circular and Sustainable) Extended Partnership and received funding from the European Union Next-Generation EU (PIANO NAZIONALE DI RIPRESA E RESILIENZA (PNRR) – MISSIONE 4 COMPONENTE 2, INVESTIMENTO 1.3 – D.D. 1551.11-10-2022, PE00000004) and  the Italian Ministry of Research, under the complementary actions to the NRRP “D34Health - Digital Driven Diagnostics, prognostics and therapeutics for sustainable Health care” Grant (\# PNC0000001) and by the PNRR Project: Silicon Valley 8.0, M1C2 - CDS000965 (CUP
C17J23000030001). She  also gratefully acknowledges the support of Gruppo Nazionale per la Fisica Matematica (GNFM) of Istituto Nazionale di Alta Matematica (INdAM) through the “INdAM - GNFM Project", codice CUP \#E5324001950001\# and of Università di Pavia for her three-month doctoral visit at the University of Galway,  Ireland.

\appendix

%%%%%%%%%%%%%%%%%%%%%%%
\section{Finite Element mesh generation}
\label{app:fe_pipeline}

%%%%%%%%%%%%%%%%%%%%%%

This appendix details the parametric mesh-generation procedure used to build the Finite Element models of Section~\ref{5.5} (Figure~\ref{fig:fea_mesh}).

In the first step, the control parameters of Table~\ref{table:fea_parameters} are used to define 2D boundary curves for a repeating quarter of a unit cell of the auxetic layer (darker shading, top left of Figure~\ref{fig:fea_mesh}); at this step, the boundary curves define the square boundary as well as the corner ellipse features.

The quarter segment's boundary regions are then filled with triangles using constrained Delaunay triangulation \cite{vandenheuvel_delaunaytriangulationjl_2024}, as implemented in the Comodo library, with node spacing $s_1$ used for both curve definition and triangulation. Both the auxetic layer and the interior are meshed in this step. Once triangulated, the quarter segment is mirrored and translated to produce the complete 2D unit cell, which is then copied in the axial and circumferential directions to produce the desired number of repetitions.

The copied 2D mesh is merged, so that nodes are shared across copies, and warped to define the outer surface of the cylinder domain; the cylinder's side and bottom surfaces are defined in the same node-sharing fashion. The two side surfaces are constructed as copies of each other, so that periodic boundary conditions can later be applied there. Node spacing is $s_1$ where nodes are shared with the outer cylinder surface, and the coarser $s_2$ everywhere else.

The cylinder's interior is then filled with tetrahedral elements, using Comodo's implementation of TetGen \cite{si_tetgen_2015}; deeper tetrahedral regions are allowed to be coarser than at the boundary, up to a factor $f$ times the surface volume. The auxetic coating is finally formed by thickening the outer surface by a thickness $t$ with node spacing $s_1$, extruding the surface triangles into pentahedral elements.

This procedure produces a fully parameterised, multi-region mesh with continuous node sharing across regions, compatible with the periodic boundary conditions described in Section~\ref{5.5}. Node, surface and element sets required for the boundary condition specification (e.g.\ the bottom, front and back surface sets shown in Figure~\ref{fig:fea_bc}) are obtained automatically as part of this process. Once the mesh and boundary condition sets are defined, the pipeline produces an ABAQUS.INP input file containing both the mesh and the boundary condition specifications. The entire process --- meshing, visualisation and INP file creation --- is efficient, typically taking about 2 seconds on a laptop computer (32 GB RAM, Intel Core Ultra 7 155H CPU).

%%%%%%%%%%%%%%%%%%%
\section{Experimental methods}
\label{app:experiments}

%%%%%%%%%%%%%%%%%%%

The specimens were fabricated using a Stratasys J35 Pro 3D printer, which employs PolyJet technology based on the layer
by layer deposition of various UV light curable polymer materials.

The sample geometries were designed in SolidWorks (Corporation, 2022).
The resulting CAD files were imported in GrabCAD Print (Stratasys, 2026) for the orientation setup, material selection and then sent to the J35 device for production.

The printer works with several material cartridges, including model and support ones, and it can mix model materials of the selected ratios automatically to produce digital materials with modified properties.
The printing time of a given sample depends on its size, number of
layers and orientation. Post processing of printed samples was performed up to 24 h after the printing
session and included the removal of the support material SUP710. 

The central cylinder was printed from 
ElasticoClear (FLX934) material, with a tensile strength of 3-5 MPa, elongation at break of 360-400\%. The grips were printed from VeroUltraWhite (RGD824), with a tensile strength of 50-65 MPa, elongation at break of 5-20\%. The meshed coating was printed from FLXA9770-DM, a digital material composed of ElasticoClear and VeroUltraWhite; its effective Poisson's ratio is estimated in Appendix~\ref{app:poisson}.

To assess the mechanical response of the samples, we performed uniaxial tensile tests on a
universal testing machine Instron 34SC-5. The specimens were clamped at the top and bottom
ends, with the bottom clamp fixed and the top clamp moving at a constant speed of 0.03 mm/s. During
the test, video and photographic recordings were made, where changes in the model were noted and
recorded according to the test time.

\begin{figure}[h]
    \centering
    \includegraphics[width=0.65\linewidth]{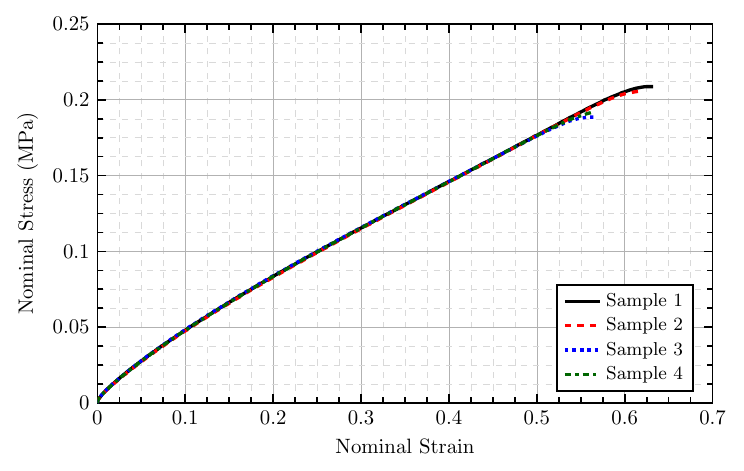}
    \caption{Comparison of nominal stress versus nominal strain for a bilayer cylindrical specimen. The substrate is modeled as incompressible ElasticoClear, and the surface film is Grade-6 (a blend of ElasticoClear and VeroUltraWhite). Results are shown for four independently tested samples.}
    \label{fig:stressvstrain}
\end{figure}

%%%%%%%%%%%%%%%%%%%%%%%%
\section{Estimation of effective Poisson’s ratio for an ElasticoClear-VeroUltraWhite hybrid}
\label{app:poisson}

%%%%%%%%%%%%%%%%%%%%%%%%%%

The digital material FLXA9770-DM (`Grade-6'), used for the meshed coating, is a mechanical blend of ElasticoClear (EC) and VeroUltraWhite (VUW) in weight fractions of 78.6\%/21.4\%, respectively, as reported by Abu-Qbeitah et al \cite{suhib2025experimental}. We estimate its effective Poisson's ratio $\nu_{\text{hybrid}}$ from a Hill average of the Voigt (iso-strain) and Reuss (iso-stress) bounds on the bulk and shear moduli of the blend.
Table~\ref{tab:hillinput} lists the mass fraction $w_i$, density $\rho_i$, Young's modulus $E_i$ and Poisson's ratio $\nu_i$ of each phase --- taken from \cite{suhib2025experimental} (elastic properties) and the Stratasys datasheets \cite{stratasys_veroultra, stratasys_elastico} (densities) --- together with the derived bulk and shear moduli $K_i = E_i/[3(1-2\nu_i)]$ and $G_i = E_i/[2(1+\nu_i)]$.

\begin{table}[h!]\centering
\begin{tabular}{@{}lccccc@{}}
\toprule
Material & $w_i$ (-) & $\rho_i$ ($\mathrm{g/cm^{3}}$) & $E_i$ ($\mathrm{MPa}$) & $\nu_i$ (-) & $(K_i,G_i)$ ($\mathrm{MPa}$) \\
\midrule
ElasticoClear (EC) & 0.786 & 1.20 & 0.16 & 0.499 & $(26.67,\ 0.053)$ \\
VeroUltraWhite (VUW) & 0.214 & 1.21 & 638 & 0.39 & $(966.67,\ 229.50)$ \\
\bottomrule
\end{tabular}
\caption{Material parameters for the two constituent phases and their derived bulk and shear moduli.}
\label{tab:hillinput}
\end{table}

Converting weight to volume fractions, $f_i = (w_i/\rho_i)/\sum_j (w_j/\rho_j)$, gives $f_{\text{EC}} = 0.787$ and $f_{\text{VUW}} = 0.213$. The Voigt and Reuss bounds,
\begin{equation}
K_V = \textstyle\sum_i f_i K_i, \quad G_V = \sum_i f_i G_i, \qquad\qquad
K_R = \Big(\sum_i f_i/K_i\Big)^{-1}, \quad G_R = \Big(\sum_i f_i/G_i\Big)^{-1},
\end{equation}
evaluate to $(K_V,G_V) = (226.5,\ 48.8)~\text{MPa}$ and $(K_R,G_R) = (33.6,\ 0.067)~\text{MPa}$, so that the Hill (arithmetic mean) estimate is $(K_H,G_H) = (130.1,\ 24.5)~\text{MPa}$.
Substituting these into $\nu = (3K-2G)/[2(3K+G)]$ gives the Voigt, Reuss and Hill estimates of the effective Poisson's ratio as $\nu_V = 0.399$, $\nu_R = 0.499$ and
\begin{equation}
{\nu_{\text{hybrid}} = \nu_H \approx 0.41,}
\end{equation}
respectively, consistent with the dominance of the near-incompressible ElasticoClear phase in the volumetric response and of the stiff VeroUltraWhite phase in shear.

\printbibliography
\end{document}